\documentclass[pss,fleqn]{w-art} 
\usepackage{times}
\usepackage{w-thm}
\usepackage{amssymb}
\usepackage{pifont}
\usepackage{graphicx}

\newcommand{\boldit}[1]{\ensuremath{\text{\bfseries\itshape #1}}}
\newcommand{\val}[2]{\ensuremath{\text{#1}\,\text{#2}}}
\newcommand{\eval}[2]{\ensuremath{#1\,\mathrm{#2}}}
\newcommand{\scival}[3]{\ensuremath{\text{#1}\times\text{10}^\text{#2}\,\text{#3}}}
\newcommand{\escival}[3]{\ensuremath{#1\times10^{#2}\,\mathrm{#3}}}
\newcommand{\magl}{\ensuremath{\ell}}
\newcommand{\nk}{_{nk}}
\newcommand{\effm}{\ensuremath{{m^*}}}
\newcommand{\eps}{\varepsilon}
\newcommand{\cyclfreq}{\ensuremath{{\omega_c}}}
\newcommand{\Ugate}{\ensuremath{U_g}}
\newcommand{\Vbias}{\ensuremath{V_\text{bias}}}
\newcommand{\subsqr}{{\scriptscriptstyle\square}}
\newcommand{\nzwd}{\ensuremath{n_\subsqr}} 
\newcommand{\percent}{\ensuremath{\,\%}}
\newcommand{\nplus}{$\text{\emph n}^+$}
\newcommand{\epsoben}{\ensuremath{\eps_\text{1}^{\protect\raisebox{.1ex}[0pt][0pt]%
{$\scriptscriptstyle\bigtriangleup$}}}}
\newcommand{\epsunten}{\ensuremath{\eps_\text{2}^{\scriptscriptstyle\bigtriangledown}}}
\makeatletter
\def\etal.{\penalty\@highpenalty\ \emph{et\penalty\@lowpenalty\ al.}}
\makeatother 
\newcommand{\micro}{{\small\Pisymbol{psy}{109}}}
\DeclareMathSymbol{\hbar}{\mathord}{AMSb}{"7D}

\begin{document}
\renewcommand{\copyrightyear}{2007}
\DOIsuffix{theDOIsuffix}
\Volume{XX}
\Issue{1}
\Month{01}
\Year{2007}
\pagespan{1}{}
\Receiveddate{23 September 2007}
\Reviseddate{7 December 2007}
\Accepteddate{10 December 2007}

\hbox{}




\title{Laterally coupled and field-induced quantum Hall systems}


\author{M. Habl\footnote{Corresponding
     author: e-mail: {\sf matthias.habl@physik.uni-r.de},
     Phone: +49\,941\,943\,2065,
     Fax: +49\,941\,943\,4226}}

     \address{Institut f\"ur Experimentelle und Angewandte Physik,
     Universit\"at Regensburg, 93040 Regensburg, Germany} 
\author{W. Wegscheider}
\begin{abstract}
A quantum Hall system which is divided into two  laterally
coupled subsystems by means of a tunneling barrier exhibits a complex
Landau level dispersion.  Magnetotunneling spectroscopy is employed to
investigate the small energy gaps which separate subsequent Landau
bands. The control on the Fermi level permits to trace the
anticrossings for varying magnetic fields.  The band structure
calculation predicts a magnetic shift of the band gaps on the scale of
the cyclotron energy. This effect is confirmed experimentally by a
displacement of the conductance peaks on the axis of the filling
factor. Tunneling centers within the barrier are responsible for
quantum interferences between opposite edge channels.  Due to the
disorder potential, the corresponding Aharonov-Bohm interferometers
generate additional long-period and irregular conductance features.
In the regime of strong localization, conductance fluctuations
occur at small magnetic fields before the onset of the regular Landau
oscillations. The Landau dispersion is obtained by a dedicated
algorithm which solves the Schr\"odinger equation exactly for a single
electron residing in a quantum Hall system with an arbitrary
unidirectional, threefold staircase potential.
\end{abstract}

\maketitle





\section{Introduction}
In the presence of a perpendicular magnetic field, the quasi-free
charge carriers of a two-dimensional electron system (2DES) condense
in equidistant and numerously degenerated Landau levels
\cite{Lan30}. Variations of the electrostatic potential locally shift
the degeneracy and, therefore, cause the formation of Landau
bands. The bending of the Landau dispersion at the edge of a 2DES is
discussed thoroughly in the literature \cite{Hal82,Mac84} as it is the
key for the understanding of the quantum Hall effect \cite{Kli80}.  If
the lateral extension of a 2DES is limited electrostatically or by the
physical edge of the underlying (lithographically defined) structure,
the Landau bands rise on a length scale which substantially exceeds
the magnetic length. However, if the confining barrier is fabricated
by means of epitaxy, namely, by the technique of cleaved-edge
overgrowth \cite{Pfe90}, the realization of a sharp edge potential
becomes possible. The corresponding edge channels are then located at
distances on the order of the magnetic length \cite{Gra04aHub05}.

In this paper, we report on the magnetotunneling spectroscopy between
two quantum Hall systems which are separated by an atomically precise
potential barrier within a GaAs/AlGaAs heterostructure. The width of
the barrier amounts to \val{52}\AA, i.e.\ it falls always below the
magnetic length. In the vicinity of the barrier, a complex Landau band
structure exists which can be described in the case of weak coupling
as a superposition of the mirror-inverted dispersions of both
subsystems.  The degeneracy at the crossings is lifted by small Landau
band gaps \cite{Ho94}.  The tunneling current through the structure
becomes maximum when the Fermi level coincides with one of these
anticrossings \cite{Kan00a}. The conductance is additionally
determined by random tunneling centers in the barrier which are
responsible for quantum interferences between opposite edge
channels \cite{Kim03aKim03b, Yan05}.

\begin{vchfigure}[t]
  \includegraphics{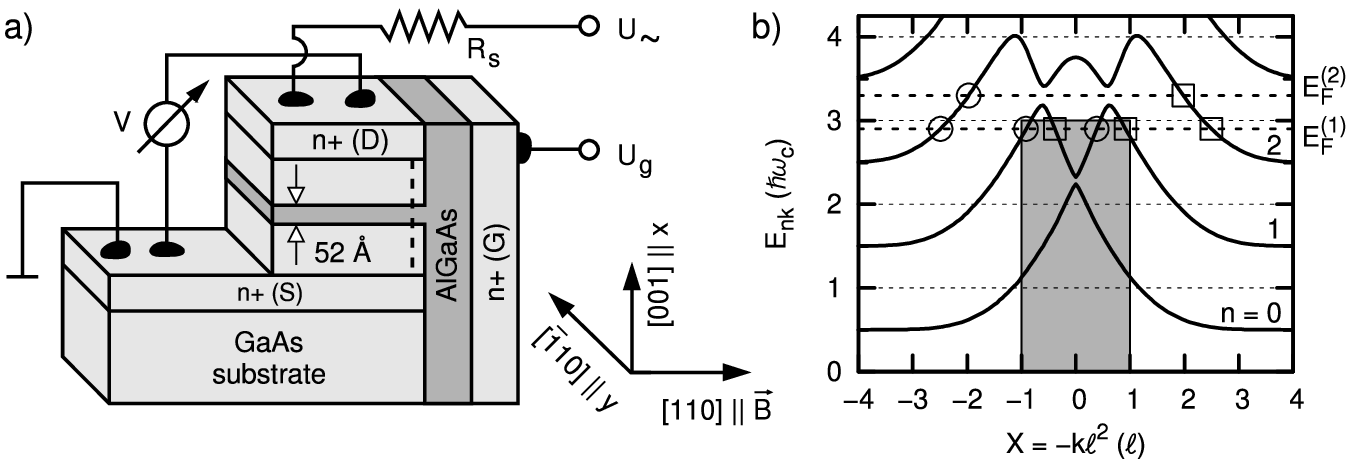}\vchcaption{\label{fig:intro}a)
  Sample design (not to scale) and measurement setup. b)~Landau
  dispersion for two strongly coupled electron systems. Edge channels
  marked with a circle (square) belong to the left (right) 2DES and
  contain electrons which propagate in the negative (positive)
  $y$-direction.}
\end{vchfigure}

While the electron systems of our structure are field-induced by means
of a gate electrode, cf. Fig.~\ref{fig:intro}a, a former experiment by
Kang\etal.\ is based on modulation-doped electron films
\cite{Kan00a}. A fixed electron density implies that a certain Landau
band gap coincides with the Fermi level only for a small range of the
magnetic field. In contrast, a sample design with a gate electrode
allows to investigate a particular band gap at different magnetic
fields provided that the Fermi level is adjusted in a suitable way.
In order to probe the same anticrossing for different barrier widths,
it is, therefore, not necessary to prepare several samples as the
effective shape of the barrier (in units of the magnetic
length~$\magl$ and the cyclotron energy~$\hbar\omega_c$) can simply be
varied by the magnetic field. Carrying out measurements for different
electron densities with the same sample is also indispensable if
quantum interferences are investigated which depend on the random
configuration of tunneling centers within the barrier. Another effect
can be observed just with a single sample, namely, the small magnetic
shift of the Landau band gaps on the scale of the cyclotron energy
which would otherwise be perturbed by the unavoidable fluctuations
between successive growth processes.

All effects discussed in this paper depend basically on the shape of
the Landau band structure at the tunneling barrier
(Fig.~\ref{fig:intro}b). For weakly coupled electron systems, Ho has
calculated the energy dispersion by using an approximation approach
\cite{Ho94}. Though the result reflects the actual  dispersion
very well, the gap positions differ noticeably from the findings of a
complete quantum mechanical calculation if a fine energy
resolution is relevant as for our experiment \cite{Hab06a,Hab06b}.
Instead of combining the dispersions of both subsystems, Takagaki and
Ploog have developed a tight-binding model which is not only
applicable for weakly, but also for strongly coupled electron systems
\cite{Tak00}. The tunneling barrier is thereby represented by a reduced
hopping amplitude between two simulation grid lines.  Since the
effective shape of the barrier is not explicitly considered in this
model, the latter is inappropriate to make accurate predictions on the
gap position in dependence of the magnetic field. 

The Landau dispersion which is necessary for the interpretation of the
obtained experimental data has been calculated exactly
by solving the single-electron Schr\"odinger equation. Since the
potential term in the Hamiltonian is given by a superposition of the
parabolic magnetic confinement potential and the piecewise constant
conduction band offset of the intrinsic semiconductor heterostructure,
the problem is generally closely related to the linear quantum
harmonic oscillator. The analytic solution becomes possible with an
ansatz for the wave function which is composed of parabolic cylinder
functions. For calculating the energy eigenstates, we have developed a
dedicated algorithm which numerically solves the continuity conditions
at the heterojunctions.  It is generally applicable to all systems
which consist of up to three regions of constant potential. Hence,
this method yields the Landau dispersion for systems with either a
rectangular barrier (biased or not) or a potential well (quantum wire)
\cite{HablCPC}.

This article is organized as follows: The sample structure and the
corresponding Landau dispersion are introduced in the next two
sections. The subject of Sec.~\ref{sec:tun} consists of two different
tunneling models which are the basis for the understanding of the
Landau oscillations and simultaneously occurring conductance
fluctuations. In Sec.~\ref{sec:basic} we discuss a method for
determining the electron density from magnetotunneling measurements.
The comparison of the conductance traces with the Landau band
structure is carried out in Sec.~\ref{sec:LL-osci} where the focus
lies on the magnetic shift of the anticrossings.  Section~\ref{sec:AB}
deals with Aharonov-Bohm oscillations at the first conductance peak on
the scale of the filling factor. Conductance fluctuations which appear
at low magnetic fields are finally discussed in
Sec.~\ref{sec:fluct_low}.

\section{Experiment}\label{sec:exp} 
For a quantitative investigation of the energy dispersion in
dependence of the magnetic field, it is necessary to have both a
well-defined tunneling barrier and electron systems of variable
density.  The molecular beam epitaxy (MBE) allows the fabrication of
GaAs/AlGaAs interfaces with a roughness which is as low as one
monolayer. For low-dimensional quantum systems like ours, one commonly
needs a sharp potential modulation not just in one, but in two
directions of space.  For this purpose, Pfeiffer\etal.\ have developed
the cleaved-edge overgrowth (CEO) method which allows to grow
subsequently two layer sequences at right angles to another
\cite{Pfe90,Sch04}.  With this technique, the sample structure of
Fig.~\ref{fig:intro}a can be realized.  It contains two laterally
adjacent electron systems which reside in the layers of the first
growth step and are induced by means of a gate electrode which is
fabricated during the second MBE step.

The source and drain contacts of the heterostructure consist of GaAs
layers which are silicon-doped to \scival2{17}{cm$^{-\text3}$} with
thicknesses of 500 and 1100$\,$nm, respectively. The quantum region is
composed of a \val{52}{\AA} thick
$\text{Al}_\text{0.34}\text{Ga}_\text{0.66}\text{As}$ barrier and two
embedding layers of intrinsic GaAs (each \val2{\micro m}). After the
first growth step, the wafer is discharged from the MBE machine and
chemically polished to a thickness of 80 to \val{100}{\micro m} by
means of a bromine methanol solution. The wafer is then cleaved into
small pieces which are scratched at a certain position and transferred
into the growth chamber again. Immediately after breaking the samples
\emph{in-situ}, the freshly exposed cleavage planes are overgrown with
\val{100}{nm} of $\text{Al}_\text{0.31}\text{Ga}_\text{0.69}
\text{As}$. The barrier region is followed by a \val{200}{nm} thick
layer of \nplus-GaAs with an electron density of
\scival2{18}{cm$^{-\text3}$}.  It acts as the gate electrode. By
etching a mesa, the buried source contact is exposed. The samples are
divided into stripes where each piece contains a \val{500}{\micro m}
long section of the cleavage plane.  Ohmic contacts on the
\nplus-layers are realized with indium droplets. These are deposited
by a soldering iron and subsequently alloyed into the crystal at about
360$^\circ$C.

If the active region of a CEO sample is accessed via \nplus-layers of
the (001) growth step, the occurrence of bulk leakage currents is
generally possible \cite{Fei06Her07}.  In the case of our structure,
the leakage current flows apart from the coupled electron systems
through the \val4{\micro m} thick intrinsic region.  In the (001)
cross-section of the heterostructure, the induced 2DESs account only
for about 1/100,000 of the total area. In order to prevent that
conductance oscillations in the quantum region cannot be measured
against the background of the bulk leakage current, the aluminum
content in the tunneling barrier has to be sufficiently high. Thus the
sample design of Fig.~\ref{fig:intro}a allows only Landau band gaps
which are considerably smaller than the cyclotron energy. Since for
such a dispersion the number of opposite edge states with significant
overlap is small in comparison to the low-barrier case which is
exemplarily shown in Fig.~\ref{fig:intro}b, this coupling regime is
denoted here as weak---independently from the absolute tunneling
resistance.

The quality of the indium contacts and of the tunneling barrier is
revealed by the two-point I-V-curves of Fig.~\ref{fig:curves}a: Both
pairs of junctions on the source and drain contact layers show a
perfect ohmic behavior. The ratio of their reciprocal resistance
coincides with the thickness ratio of the underlying \nplus-contact
layers.  According to the $I$-$V$ curve D$\to$S, which represents the
leakage current between source and drain for a vanishing gate voltage,
the bulk resistance amounts at least \val{21}{k$\Omega$} for applied
voltages below~\val{10}{mV}. Therefore, magnetotransport measurements
can be carried out with AC voltages which are small enough so that the
bulk resistance always exceeds the resistance of the induced electron
systems.

\begin{vchfigure}[t]
  \includegraphics{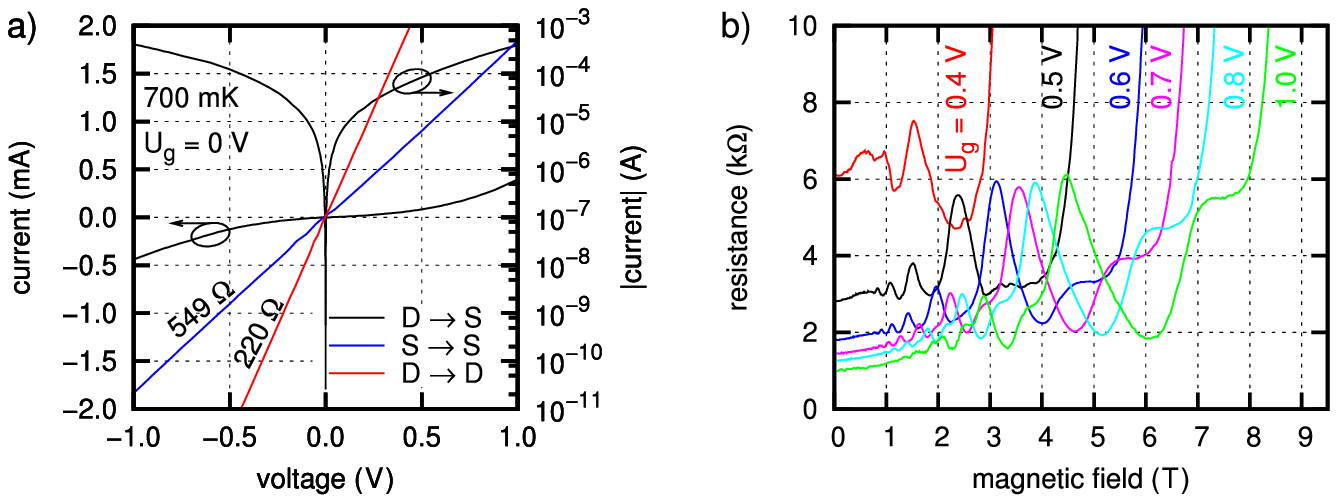}\vchcaption{\label{fig:curves}a)
  Two-point current-voltage ($I$-$V$) traces for three different
  combinations of the source~(S) and drain~(D) contacts.  The bulk
  leakage current D$\to$S (black curves) is shown both on a linear and
  logarithmic scale. b)~Magnetotunneling measurements for different
  gate voltages at \val{400}{mK}.}
\end{vchfigure}

The magnetotransport properties of the quantum region have been
investigated by means of lock-in technique. The corresponding circuit
is shown in Fig.~\ref{fig:intro}a. In spite of the four-point contact
geometry, the active region is effectively accessible only via two
leads, namely, the extended \nplus-layers of the first growth step.
This fact prevents a direct determination of the charge carrier
density by standard methods like Shubnikov-de Haas or (quantum) Hall
measurements.  The coupled electron systems have been studied by
applying an AC voltage of $U_\sim=\val1V$ via a high-impedance
resistor $R_s=\val{100}{M$\Omega$}$. This setup ensures a practically
constant current of $I=(U_\sim-V)/R_s\simeq\val{10}{nA}$.  A
quasistatic measurement is enabled by a low AC frequency of
\val{17}{Hz}.  All measurements have been carried out within a
$^\text{3}$He system which provides temperatures down
to~\val{350}{mK}.

The resistance traces shown in Fig.~\ref{fig:curves}b in dependence of
the magnetic field bear some resemblance to the Shubnikov-de Haas
effect. In the following, the prominent long-period resistance
oscillations are called \emph{Landau oscillations} as they originate
from the periodic coincidence of the Fermi level with the Landau band
gaps. Other magnetotransport features are two different types of
conductance fluctuations which appear respectively at low field
strengths ($B\lesssim\val1T$) and in the range of the rightmost
resistance minimum (in Fig.~\ref{fig:curves}b best observable for
$U_g=\val{0.5}V$). For the understanding of all phenomena, it is
necessary to know the Landau band structure at the tunneling barrier
quantitatively.

\section{Landau band structure}\label{sec:disp}
Provided that the electrostatic potential of a two-dimensional Hall
system is constant, the energy levels of the charge carriers are given
by the solutions of the quantum harmonic oscillator \cite{Lan30}.
Close to potential variations, the degeneracy of the Landau levels is
lifted so that they form dispersive bands.  The particular Landau
bands increase monotonically for electrons approaching the confinement
potential of a 2DES \cite{Mac84, Bar94}.  In the vicinity of a
tunneling barrier, the Landau dispersion shows an oscillating
structure which is calculated in the following. With regard to the
conduction band shape of the GaAs/AlGaAs/GaAs heterostructure, the
treatment is restricted to electron systems which are unidirectionally
modulated by three regions of constant potential. The uniform
potential within each region makes a representation of the energy
eigenfunctions by parabolic cylinder functions possible. Besides, the
constraint concerning the number of heterojunctions keeps the equation
system which results from the continuity conditions solvable.  For a
system with a single potential discontinuity, one has to determine
only the energy eigenvalue which can be done by means of a standard
solution method. A sample structure with two heterojunctions already
requires a dedicated solution algorithm for the simultaneous
calculation of both the energy eigenvalue and a characteristic mixing
parameter.  For the analytic and numerical details we refer to
Refs.\ \cite{Hab06b} and~\cite{HablCPC}.

The Schr\"odinger equation which describes the whole quantum region is
solved in the scope of the effective mass approximation. For the GaAs
layers and the AlGaAs barrier, we use the same effective mass
$\effm=\text{0.067}m_e$.  The aluminum content $x=\text{0.31}$ of the
potential barrier corresponds to a conduction band offset of
$V_0=\val{268}{meV}$ with respect to GaAs \cite{Ada94}. As the barrier
height is a multiple of the cyclotron energy
$\hbar\omega_c\approx\val{1.73}{meV}\times \left(B/\val1T\right)$, the
probability density for electrons at typical magnetic fields is very
low inside the AlGaAs material.  Consequently, the assumption of a
constant effective mass throughout the whole structure is a rather
good approximation. For strongly coupled electron systems, however,
the difference concerning both the cyclotron energy and the cyclotron
radius becomes absolutely relevant~\cite{HablCPC}.

If the conduction band offset is assumed to vary just along the
$x$-axis, i.e.\ in the direction of the first growth step, the
Schr\"odinger equation for a single electron subjected to a magnetic
field reads
\begin{equation}
  \left\{\frac1{2m^*}\left(\mspace{3mu}\boldit p+
    e\mspace{1mu}\boldit A\right)^2+V(x)\right\}
  \psi(x,y)=E\psi(x,y).
  \label{eqn:SGl2dim}
\end{equation} 
For electrons propagating parallel to the tunneling barrier and
perpendicular to $\boldit B=(0,0,B)$, it is appropriate to use the
Landau gauge $\boldit A=\left(\text0,xB,\text0\right)$.  This
definition of the vector potential implies a wave function which
represents a localized state in the $x$- and a plane wave in the
$y$-direction:
\begin{equation}
  \psi_{nk}(x,y)=\frac1{\sqrt{L_y}}e^{iky}\varphi_{nk}(x).
  \label{eqn:ansatz}
\end{equation}
The integer $n=\text0$, 1, 2,~$\ldots$ is the Landau band index, and
$k$~stands for the angular wavenumber which determines the wavelength
along the $y$-direction.  The extension~$L_y$ of the system is
contained in the ansatz of Eq.~\eqref{eqn:ansatz} for the purpose of
normalization.  The $x$-dependent component $\varphi_{nk}$ describes a
bound state.  By using Eq.~\eqref{eqn:ansatz} and the magnetic length
$\ell=\sqrt{\hbar/eB}$, the Schr\"odinger equation transforms to
\begin{equation}
    \left\{\frac{\hbar^2}{2\effm}\left[\left(k+\frac x{\magl^2}\right)^2
  -\frac{d^2}{dx^2}\right]+V(x)\right\}\varphi\nk(x)=E\nk\varphi\nk(x).
  \label{eqn:SGl1dim}
\end{equation}
The elimination of the $y$-coordinate reduces the partial differential
equation~\eqref{eqn:SGl2dim} to an ordinary differential equation
which now solely determines the localized state~$\varphi\nk(x)$. 

The $x$-quadratic term of the Hamiltonian represents the magnetic
confinement potential which is given by a parabola centered at
$X=-k\magl^2$. Only for bulk states, which are not influenced by
potential variations, the quantity~$X$ is always identical to the
quantum mechanical expectation value of the electron location. If $X$
resides, e.g., within an infinitely high potential step, the
probability density vanishes at~$X$ and the center of mass deviates
from this location.  In the semiclassical picture, the variable~$X$
stands for the center of the skipping orbits of electrons propagating
along a barrier.  Therefore, the parameter~$X$ is commonly called
guiding center. While dispersion relations are generally plotted
against the wavenumber~$k$, the close connection between the guiding
center and the actual location of the electron often gives reasons to
use $X=-k\magl^2$ as the abscissa in plots of the Landau band
structure. This enables a direct comparison between the dispersion and
the potential landscape even when $X$ substantially differs from the
actual electron position.

\begin{vchfigure}[t]
  \includegraphics{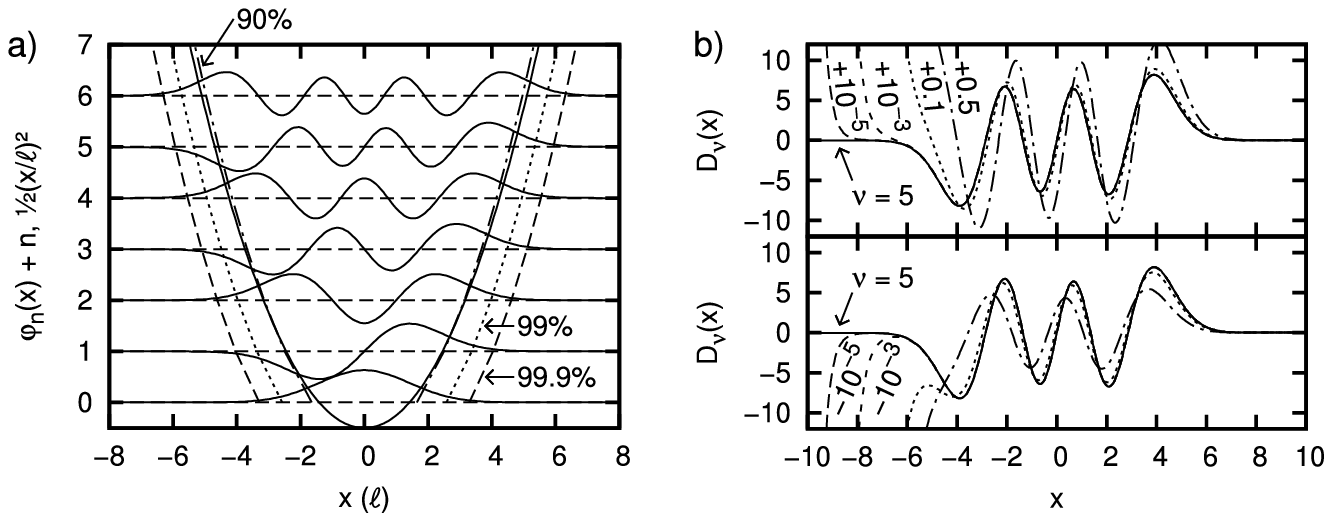}\vchcaption{\label{fig:hermite}a)
  Normalized wave functions for bulk states with $X=0$. The solid
  parabola represents the magnetic confinement potential in units
  of~$\hbar\cyclfreq$. The broken lines enclose regions with
  $\int_{-x}^{+x}\varphi_n^2(x^\prime)\,dx^\prime=\text{0.90}$, 0.99,
  and 0.999, respectively. b)~The graph illustrates the behavior of
  $D_\nu(x)$ for a~$\nu$ which approaches an integer from higher or
  lower values. The signed numbers denote the difference to
  $\nu=\text5$.}
\end{vchfigure}

Electrons which occupy bulk states are described by wave
functions~$\varphi\nk(x)$ which are already converged to zero at the
next potential variation. For these electrons, the term~$V(x)$ may be
neglected in Eq.~\eqref{eqn:SGl1dim} which then becomes identical to
the Schr\"odinger equation for a quantum harmonic oscillator.
Therefore, in regions where the conduction band is flat, the potential
in the Hamiltonian is---except for a constant---completely determined
by the effect of the Lorentz force:
\begin{equation}
  V_B(x)=\frac{\hbar^2}{2\effm}\left(k+\frac x{\magl^2}\right)^2=
  \frac12\frac{(x-X)^2}{\magl^2}\hbar\cyclfreq=\frac12\xi^2\hbar\cyclfreq.
\end{equation}
The last term contains with $\xi\equiv(x-X)/\magl$ a dimensionless
variable which is used in the following in place of~$x$.  Bulk
electrons which are bound within the parabolic confinement
potential~$V_B(x)$ occupy the equidistant energy levels
$E_n=\left(n+\frac12\right)\hbar\cyclfreq$. The corresponding wave
functions are given by the Hermite functions: $\varphi_n(x)\propto
e^{-\xi^2/2}H_n(\xi)$. From Fig.~\ref{fig:hermite}a it follows that
the extension of the wave functions is approximately limited by the
parabola $\xi^2/2=V_B(x)/\hbar\cyclfreq$. In particular, about
90\percent{} of all electrons reside within the interval
$[-\xi^2/2;\xi^2/2]$. Hence, one can deduce a cyclotron radius of
$R_c=\sqrt{2n+1}\magl$ which is in fact the half extension of the wave
function. For electrons with a guiding center which is located at a
distance of about $R_c$ or less in respect of a potential step, the
quantum harmonic oscillator is no longer a good approximation.

For solving the Schr\"odinger equation~\eqref{eqn:SGl1dim} separately
for each region of constant potential, it is necessary to admit for a
moment the most general solution which also includes diverging
eigenfunctions. By using the dimensionless variables $\xi$ and
$\eps\nk$, where the latter is defined according to
$E\nk\equiv\left(\eps\nk+ \frac12\right)\hbar\cyclfreq$, the
Schr\"odinger equation transforms to the Weber differential
equation. For the $i$-th region of constant potential
$v_i=V_i/\hbar\cyclfreq$, we get
\begin{equation}
  \left\{\frac{d^2}{d\xi^2}-\xi^2+2\left[\Bigl(\eps\nk-v_i\Bigr)
  +\frac12\right]\right\}\varphi\nk(\xi)=0.
  \label{eqn:WeberDEq}
\end{equation}
The corresponding two-dimensional solution space is
spanned by the linearly independent parabolic cylinder functions
$D_{\eps\nk-v_i}(\pm\xi\sqrt2)$ \cite{Erd53Spa87}.  If both outermost
intervals are of the same material---in our case GaAs---, it is
appropriate to set $v_0=v_2=0$ which leads to the total wave function
\begin{equation}
  \varphi\nk(\xi)=\gamma\begin{cases}
    \alpha_-D_{\eps\nk}(-\xi\sqrt2)&\qquad\phantom{\xi_-\le{}}\xi<\xi_-,\\
    \phantom{\alpha_-}D_{\eps\nk-v}(-\xi\sqrt2)+\beta\nk D_{\eps\nk-v}(\xi\sqrt2)
      &\qquad\xi_-\le\xi\le\xi_+,\\
    \alpha_+D_{\eps\nk}(\xi\sqrt2)
      &\qquad\xi_+<\xi.\end{cases} 
\end{equation}
The parameters $\xi_\pm=\pm\frac a{2\magl}-\frac X\magl$ denote the
positions of both heterojunctions which are assumed to be located
symmetrically around the origin. As illustrated in
Fig.~\ref{fig:hermite}b, the function $D_\nu(x)$ diverges on the
negative $x$-axis if $\nu$ is not an integer.  Outside the
barrier region, the wave functions are represented, therefore, just
by one of both basis functions. The essential parameters of the system
are the energy eigenvalue~$\eps\nk$ and the scalar $\beta\nk$. The
latter represents the mixing ratio of both parabolic cylinder functions
within the AlGaAs region.

The unknown variables $\eps\nk$ and $\beta\nk$ as well as the
prefactors~$\alpha_-$ and $\alpha_+$ are determined by the requirement
of continuity for both $\varphi\nk$ and $\varphi\nk^\prime$ at the
heterojunctions. Alternatively, one may also demand a continuous
transition of $\varphi\nk$ and $\log^\prime(\varphi\nk)$ which allows
to determine $\eps\nk$ and $\beta\nk$ independently from~$\alpha_\pm$.
The simultaneous solution of the two corresponding equations, which
are highly non-linear, has been realized by a dedicated algorithm
which is discussed in detail elsewhere \cite{Hab06a,Hab06b,HablCPC}.
For a set of (equidistant) $X$-values, the method yields (almost) all
solution tuples $(\eps\nk;\beta\nk)$ within a given energy range. The
obtained eigenstates are, however, not yet assigned to the particular
Landau bands.  For the determination of the band index~$n$, one may
exploit the properties of the discrete energy spectrum of an
one-dimensional Schr\"odinger equation: The energy levels are
non-degenerate and, if arranged according to ascending energy, the
corresponding wave functions possess 0, 1, 2,~$\ldots$ nodes
\cite{Lan65}. Thus in order to obtain the actual Landau dispersion, it
is not only necessary to solve the continuity conditions, but also to
count subsequently the zeros of the energy eigenfunctions.

\begin{vchfigure}[t]
  \includegraphics{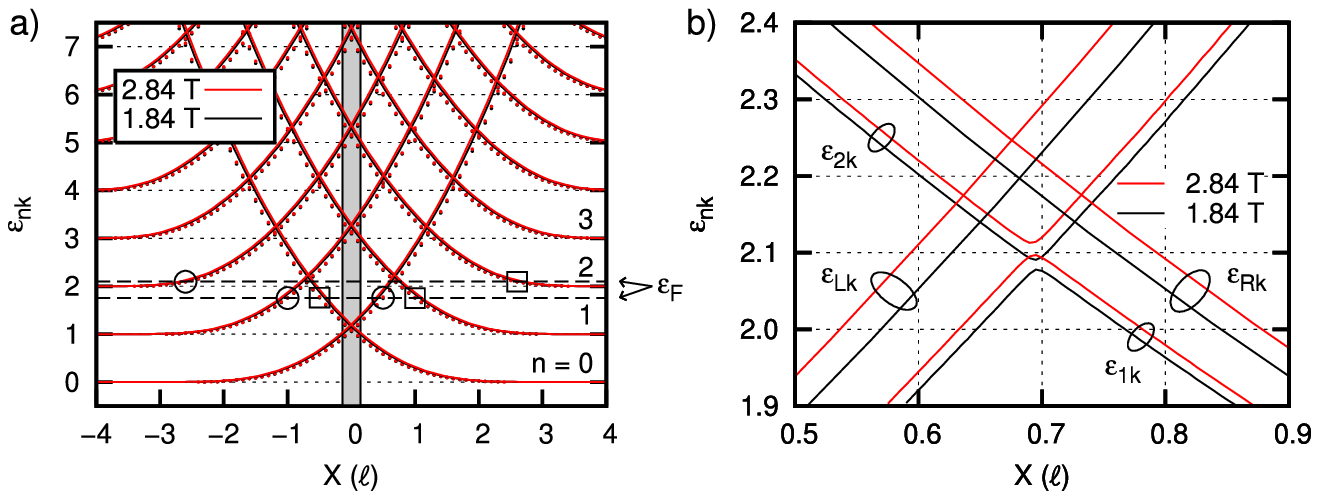}\vchcaption{\label{fig:weak}a)
  Landau dispersion of two weakly coupled electron systems which are
  separated by a \val{268}{meV} high and \val{52}{\AA} thick tunneling
  barrier. The plot compares the complete quantum mechanical
  calculation (dots) with the approximated dispersion (lines)
  according to Ref.~\cite{Ho94}.  For the Fermi levels
  $\eps_F=\text{1.75}$ and 2.1, circles (squares) mark the guiding
  centers of the edge channels in the left (right) electron system. As
  a consequence of the scaling in units of $\magl$ and
  $\hbar\cyclfreq$, the dispersions for \val{1.84}T and \val{2.84}T
  are only hardly distinguishable. b)~Details of a certain band gap:
  $\eps_{1k}$ and $\eps_{2k}$ represent the common energy eigenstates
  of both systems while the dispersion relations of two electron
  systems which are isolated by a barrier of the same width, but
  infinite height are denoted by $\eps_{Lk}$ and $\eps_{Rk}$. For an
  increase of the magnetic field of~\val1T, the anticrossing is
  shifted by $\text{0.02}\hbar\cyclfreq$.}
\end{vchfigure}

The band structure of our sample structure is depicted in
Fig.~\ref{fig:weak}a for two different magnetic field amplitudes.
When the barrier height is considerably greater than the Fermi level,
the coupling of the adjacent electron systems is weak and the shape of
the dispersion in units of $\magl$ and $\hbar\cyclfreq$ varies only
slightly in dependence of the magnetic field. A certain anticrossing
is resolved in Fig.~\ref{fig:weak}b. An increase of the magnetic field
by \val1T shifts the band gap by
$\text{0.02}\hbar\cyclfreq$. Simultaneously, the gap is broadened from
$\text{0.013}$ to $\text{0.016}\hbar\cyclfreq$.  Generally, the size
of the band gaps cannot be resolved within our experiment, but their
position is an accessible quantity. By changing the Fermi level in an
appropriate way, it is possible to trace the position of the
anticrossings while the magnetic field is varied. On large length and
energy scales, the approximation of Ho \cite{Ho94} reproduces the
actual Landau band structure very well
(Fig.~\ref{fig:weak}a). However, in comparison to the complete quantum
mechanical calculation, the position of the anticrossings is about
$\text{0.1}\hbar\cyclfreq$ higher and the increase of the gap
positions as a function of~$B$ is about twice as large as for systems
where the coupling is switched on (Fig.~\ref{fig:weak}b).

\section{Tunneling models}\label{sec:tun}
All features of the magnetotransport curves shown in Fig.~\ref{fig:curves}b
are explainable within the scope of two different models which have
been developed recently. While each describes an ideal situation, the
actual sample structure possesses essential attributes of both.  The
model of \emph{Landau level mixing} \cite{Ho94,Kan00a} is based on a
tunneling barrier which is invariant in the $y$-direction. This
property is also a prerequisite for our band structure
calculation. In contrast, the second model does not explicitly
consider the dispersion of the Landau levels. It is rather based on
the assumption that there exist \emph{tunneling centers} within the
potential barrier \cite{Kim03aKim03b,Yan05}. The model is especially
valuable for the understanding of quantum interference patterns at
certain regions of the conductance traces.

The formation of the Landau oscillations is illustrated in
Figs.~\ref{fig:tun}a and~b. The coupling of both electron systems is
determined by the degree of mixing between the edge states on the left
and right side of the tunneling barrier. The coupling becomes maximum
at the band extrema near the anticrossings. Figure~\ref{fig:weak}a
depicts the configuration of the edge channels for two different Fermi
levels. For $\eps_F=\text{1.75}$, there exist two pairs of opposite
channels containing electrons which counterpropagate along the
barrier.  When the Fermi level~$\eps_F$ increases and exceeds an
integer value, two additional edge channels emerge at all sample edges
as well as parallel to the tunneling barrier. However, if the Fermi
level increases further and enters one of the Landau band gaps, two
edge states disappear again at each anticrossing while the
corresponding channels still exist at the other edges of the coupled
quantum Hall systems. From this reason and due to the fact that the
chirality of the system suppresses backscattering into the contacts,
the electrons in the remaining parts of the edge channels are supposed
to tunnel immediately through the endings of the barrier
\cite{Kan00a}. If the Fermi level lies between the $n$-th and
$(n+1)$-th Landau band ($n=\text0$, 1, 2,~$\ldots$) and if the system
is spin-degenerate, this effect gives rise to a conductance of
$2(n+1)e^2/h$.  The authors of Ref.~\cite{Tak00} show in the scope of
a tight-binding model that the tunneling of the electrons is not strictly
limited to the outermost regions of the barrier. In fact, their
simulation demonstrates that the electrons may propagate along the
barrier for a certain distance until all of them have successively
tunneled through the potential wall.  While both anticrossings around
$\eps=2.1$ in Fig.~\ref{fig:weak}b are located at the same energy
level due to symmetry, this does generally not hold for energy gaps
between higher order bands. However, in weakly coupled electron
systems, multiple band gaps of same order overlap with each other so
that they are not distinguishable by magnetotransport measurements
\cite{Hab06b}.

\begin{vchfigure}[t]
  \includegraphics{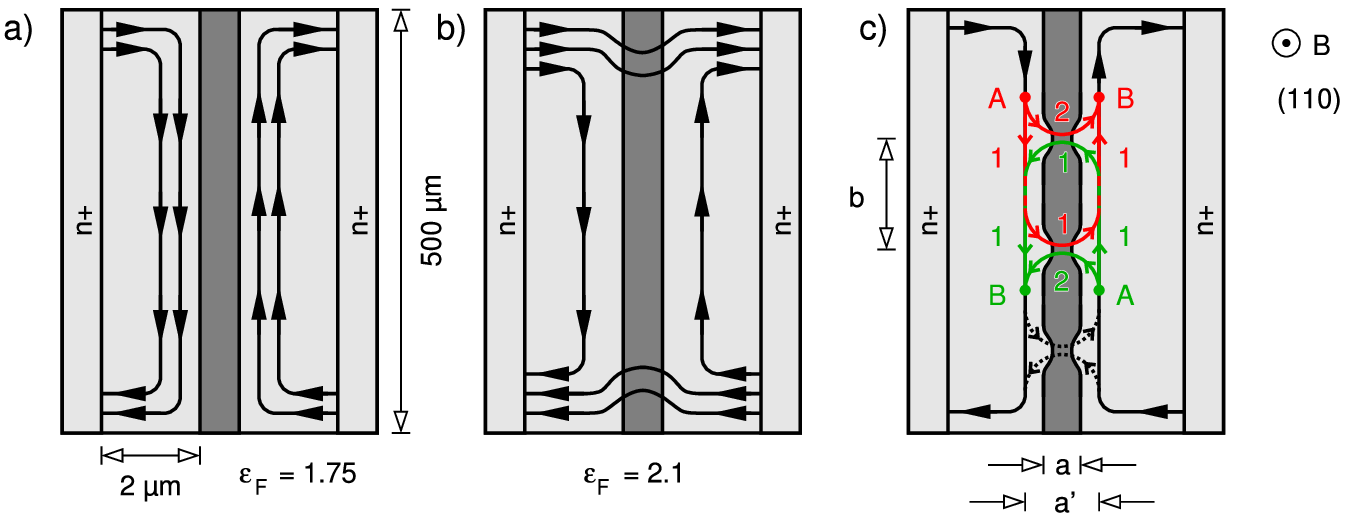}\vchcaption{\label{fig:tun}
  (110)-View of the CEO sample (not to scale). 
  a)~Configuration of edge channels for the lower Fermi level plotted in
  Fig.~\ref{fig:weak}a. The electrons propagate along the
  \val2{\micro m} long edges of the sample and parallel to the AlGaAs
  barrier. b)~Situation for a Fermi level which lies between the Landau
  bands $n=\text1$ and~2. c)~Sketch of a system with three tunneling
  centers. The red and green trajectories are
  equivalent.  An electron propagates from A to~B by either taking
  path 1 or~2. If the lower tunneling center is considered
  additionally, more complex interferences patterns become possible.}
\end{vchfigure}

The existence of tunneling centers within the barrier is the origin
for quantum interferences between counterpropagating edge channels
[9-12].  Impurities within the barrier enhance the tunneling
probability locally. The same holds for a reduced aluminum content in
the barrier which varies in the 18~monolayers thick AlGaAs layer due
to statistical fluctuations.  The illustration of Fig.~\ref{fig:tun}c
shows that for each pair of tunneling centers there exist two
different sets of trajectories which both enclose the same magnetic
flux $\Phi=Bba'$.  The Aharonov-Bohm (AB) effect is possible as an
electron at point~A may interfere with itself at point~B by taking the
paths 1 or~2 which have the lengths $2b+a'$ and $a'$, respectively.
Constructive interference takes place if the enclosed magnetic flux is
a multiple of the Dirac flux quantum $\Phi_0=h/e$. Therefore, the
period of the conductance fluctuations is given by
\begin{equation}
  \Delta B=\frac1{ba'}\frac he.  \label{eqn:AB_DeltaB}
\end{equation}
For the distance~$a'$ between the interfering edge channels we use in
the following the difference of the corresponding centers of mass on
the $x$-axis. Generally, the quantity~$a'$ can exceed the barrier
width by a multiple. In conventional AB rings, the interfering
trajectories are fixed within the current paths which are defined by
means of lithography \cite{Web85Web88}.  The course of the
trajectories in our samples is, however, not only determined by the
underlying structure, but also by the magnetic field so that
variations of the magnetic length are relevant for the enclosed
magnetic flux.  For two interfering edge states at some energy and
with guiding centers at~$\pm X$, the length~$a'$ can be estimated to
\cite{Hab06b}
\begin{equation}
  a'\approx\frac a2+|X|+R_c\approx R_c\simeq\sqrt\nu\magl
  \label{eqn:approx_distance}
\end{equation}
where the cyclotron radius is  given by
$R_c=\sqrt{2\eps\nk+1}\magl$.  The last term of
Eq.~\eqref{eqn:approx_distance} is valid for a flat density of states
as discussed in Sec.~\ref{sec:basic}. A more accurate specification
of~$a'$ becomes possible by explicitly using the results of the band
structure calculations. For the evaluation of
$\langle\varphi\nk|x|\varphi\nk\rangle$, there exists an analytic
expression for the antiderivative appearing in the corresponding
piecewise defined integral \cite{Hab06b,HablCPC}.

The interference of the electrons is perturbed by thermal
fluctuations. Depending on temperature~$T$ and the distance~$b$
between the involved tunneling centers, one can distinguish between a
coherent and incoherent regime \cite{Kim03aKim03b}.  An electron with
the velocity~$v$ takes the time $\Delta t\simeq 2b/v$ to travel along
pathway~1 outlined in Fig.~\ref{fig:tun}c. According to the
energy-time uncertainty principle, the average fluctuation of the
particle energy is then $\Delta E\ge\hbar/(2\Delta t)$. For an electron
within the coherent regime, $\Delta E$ has to be considerably greater
than the average amplitude of the thermal fluctuations. The
corresponding relation $\hbar v/(4b)\gg k_BT$ leads to the
definition of a critical temperature which specifies the transition
between the coherent and incoherent regimes:
\begin{equation}
  T_\text{AB}\equiv\frac{\hbar v}{4bk_B}=
  \frac{\hbar^2\bigl|\langle x\rangle-X\bigr|}{4bk_B\effm\magl^2}.
  \label{eqn:Tab}
\end{equation}
The last term is based on the fact that the particle momentum is given
by $p_y=\frac\hbar\magl \frac{\langle x\rangle-X}\magl$
\cite{Hab06b}. As the Landau dispersion is widely invariant in respect
of the magnetic field (if plotted against $X/\magl$ as in
Fig.~\ref{fig:weak}), the quantity $|\langle x\rangle-X|/\magl$ is
approximately constant and for the coherent regime it holds
\begin{equation}
  T\ll T_\text{AB}\propto\frac1{b\magl}.  \label{eqn:coherence}
\end{equation}
Because  a real sample likely contains not just two, but
several tunneling centers, there exist many competing interference
paths. Therefore, one expects quasiperiodic conductance fluctuations
instead of sinusoidal oscillations.  The coherence condition of
Eq.~\eqref{eqn:coherence} depends on the temperature as well as on the
magnetic field strength. Both parameters influence the contribution of
the particular tunneling centers to the conductance of the whole
system. For a decreasing temperature or a rising magnetic field, an
increasing number of different interference paths affects the
conductance. However, the $B$-dependence cannot be investigated as the
observed AB oscillations are restricted to just one conductance peak
\cite{Yan05,Hab06a}.

\section{Basic electronic properties}\label{sec:basic}
The comparison of the magnetotransport traces of
Fig.~\ref{fig:curves}b with the corresponding Landau dispersions
requires the knowledge of the exact charge carrier concentration in the coupled
electron systems.  Our sample structure does not, however, allow a
direct measurement of this quantity as each electron system is
effectively accessible just by one contact. Nevertheless, there exist
two methods which offer the possibility to determine the electron
density. On one hand, the electron concentration can be predicted by
means of a capacitor model and on the other hand it is possible to
infer it from magnetotransport measurements---similar to the
analysis of conventional Shubnikov-de Haas oscillations.

The gate element of the heterostructure in Fig.~\ref{fig:intro}a can
be modeled by a plate capacitor. The dielectric is
characterized by both the relative permittivity
$\epsilon_r=\text{11.6}$ of Al$_\text{0.31}$Ga$_\text{0.69}$As at low
temperatures \cite{Ada94} and the width $d=\val{100}{nm}$ of the
potential barrier.  From the corresponding capacitance $C=\epsilon_r\epsilon_0A/d$
it follows
\begin{equation}
  \nzwd(\Ugate)=\frac{\epsilon_r\epsilon_0}{de}\left(\Ugate-U_0\right)
  \approx\escival{6.4}{11}{cm}^{-\text2}\frac{\left(\Ugate-U_0\right)}{\eval1V}.
  \label{eqn:platecap}
\end{equation}
The voltage~$U_0$ is a sample-specific offset which is determinable
just by experiment. Because Eq.~\eqref{eqn:platecap} only predicts the
slope of $\nzwd(\Ugate)$,  a measurement method is still
indispensable for obtaining an absolute value for the electron
density.

Alternatively to the capacitor model, the electron concentration can
be inferred from the resistance traces $R(B)$ shown in
Fig.~\ref{fig:curves}b.  This method exploits the fact that the
anticrossings of the Landau dispersion in Fig.~\ref{fig:weak}a are
equidistant. Figure~\ref{fig:density}a confirms this feature by plotting 
the gap position versus the index of the particular lower Landau band.
The following two equations  reproduce the linear relation between
the band index and the energy level of the anticrossings:
\begin{subequations}\label{eqn:shift-ac}
\begin{gather}
  B=\eval{1.84}T:\qquad\eps_n^\text{ac}=(1.075\pm0.005)+(1.007\pm0.001)\,n,\\
  B=\eval{2.84}T:\qquad\eps_n^\text{ac}=(1.094\pm0.003)+(1.009\pm0.001)\,n.
\end{gather}
\end{subequations}
All energy gaps are shifted about $\Delta\eps=\text{1.08}$ with
respect to the corresponding bulk Landau levels. The collective
magnetic shift of $\text{1.094}-\text{1.075}=\text{0.019}$ reproduces
the increase $\text{0.02}\hbar\cyclfreq$ which is visible in
Fig.~\ref{fig:weak}b. The field dependence contained in the particular
second terms of Eq.~\eqref{eqn:shift-ac} is one order of magnitude
weaker than the collective shift, but it increases linearly with the
band index.  For the average field strength of~\val3T of the
conductance traces in Fig.~\ref{fig:curves}b, the distance of the
equally spaced band gaps is approximately $\Delta\eps=\text{1.01}$.
The spacing of the anticrossings is thus almost identical to the
Landau splitting of bulk states.

\begin{vchfigure}[t]
  \includegraphics{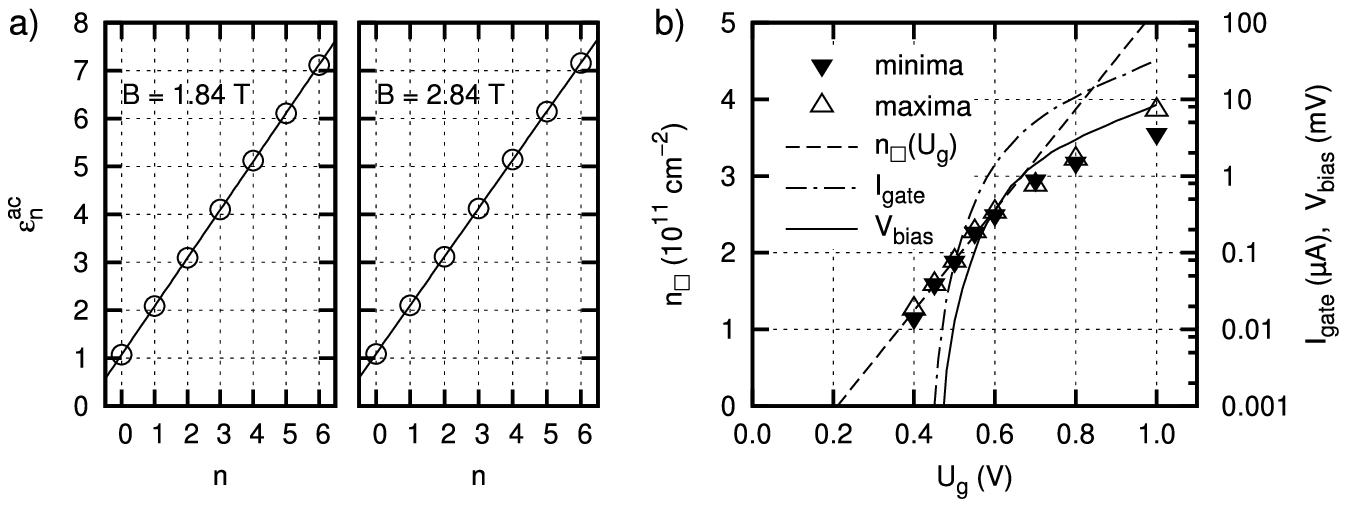}\vchcaption{\label{fig:density}a)
  Mean position of the anticrossings between the $n$-th and $(n+1)$-th
  Landau band of Fig.~\ref{fig:weak}a. The least-squares fits for the
  straight lines are reproduced by Eq.~\eqref{eqn:shift-ac}. b)
  Electron density~$\nzwd$, leakage current~$I_\text{gate}$, and
  internal bias voltage~$\Vbias$ vs gate voltage~$\Ugate$.  The data
  points represented by triangles are the result of a Shubnikov-de
  Haas analysis of $R(B)$ in conjunction with
  Eq.~\eqref{eqn:nzwd}. The dashed line which is fitted for
  $\Ugate\le\val{0.6}V$ corresponds to
  $\nzwd(\Ugate)=\escival{6.5}{11}{(\mathnormal\Ugate-\eval{0.21}V)\,cm^{-2}/V}$.}
\end{vchfigure}

The filling factor results from the electron density according to
$\nu=\nzwd h/eB$. In order to use~$\nu$ as a natural comparative
quantity for the dimensionless energy eigenvalue~$\eps\nk$, it is
still necessary to know the density of states which actually
determines the Fermi level. The authors of Refs.\ \cite{Gra04aHub05}
and~\cite{Kan00a} state an electron mobility of
${\sim}\val{10$^\text{5}$}{cm$^\text{2}$/Vs}$ for their electron films
which reside like ours along the cleavage plane of a CEO
sample. For the analysis of magnetotunneling spectroscopy experiments,
they assume a strong broadening of the Landau levels and employ a flat
density of states. The comparison between the conductance traces of
Fig.~\ref{fig:curves}b and those of Refs.\ \cite{Gra04aHub05} and~\cite{Kan00a}
makes it reasonable to use here the same asymptotic approximation
$\nu\simeq2\eps_F+1$. The distance
$\Delta\eps$ of the anticrossings then corresponds to a period of
$\Delta\nu=2\Delta\eps$ on the scale of the filling factor.  If
$\Delta(1/B)$ denotes the mean spacing of the resistance minima on the
axis of the reciprocal magnetic field, the two-dimensional electron
density can be calculated according to
\begin{equation}
    \nzwd=\frac{2e\Delta\eps}{h\,\Delta(1/B)}. \label{eqn:nzwd}
\end{equation}
A detailed error discussion in Ref.~\cite{Hab06b} yields that the
neglect of the small field dependence contained in~$\Delta\eps$
does not considerably influence the results which are presented in the
following on the basis of this relation.

Figure~\ref{fig:density}b shows the electron density obtained from the
minima and maxima of the resistance traces of
Fig.~\ref{fig:curves}b. The data points for intermediate values of the
gate voltage are gained from other measurements which are not shown
here.  The electron density increases linearly with~$\Ugate$ until the
accumulation of electrons begins to saturate at~\val{0.7}V. For gate
voltages below this value, the slope of the data points is
\scival{6.5}{11}{cm$^{-\text2}$/V} with an uncertainty of~5\percent.
Thus, the determination of the electron density according to
Eq.~\eqref{eqn:nzwd} corresponds very well to
Eq.~\eqref{eqn:platecap}, namely, to the capacitor model. This
agreement also confirms that the (enhanced) Zeeman splitting is not
resolved with our samples.

With an increasing gate voltage, at some point a notable leakage
current~$I_\text{gate}$ occurs through the AlGaAs layer of the gate
structure. For the experiment, it is not so much the leakage current
which is relevant, but rather an accompanying bias voltage which
builds up as a consequence of the current flow. Electrons are
tunneling along the gradient of the gate voltage from the electron
films into the \nplus-layer of the gate electrode. While in the lower
electron system the charge carriers are easily replaced by electrons
from the source contact (Fig.~\ref{fig:intro}a), the upper system
depletes at some degree because the electron supply from the common
ground is restricted by the \val{52}{\AA} thick tunneling
barrier. Therefore, an \emph{internal} bias voltage emerges which is
approximately proportional to the leakage current
(Fig.~\ref{fig:density}b). This bias shifts the electron systems
against each other. The consequences are the same as discussed in
Refs.\ \cite{Ho94} and~\cite{Kan00a} for an applied \emph{external}
bias voltage.  The following quantitative comparison of the measurement
results and the Landau dispersion is restricted to gate
voltages~$\le\val{0.5}V$. On that condition, a distortion of the
conductance traces due to an internal bias can be excluded as the
offset between both electron systems is then negligible small:
$e\Vbias/\hbar\cyclfreq<\text5\times\text{10}^{-\text3}$.

\begin{vchfigure}[t]
  \includegraphics{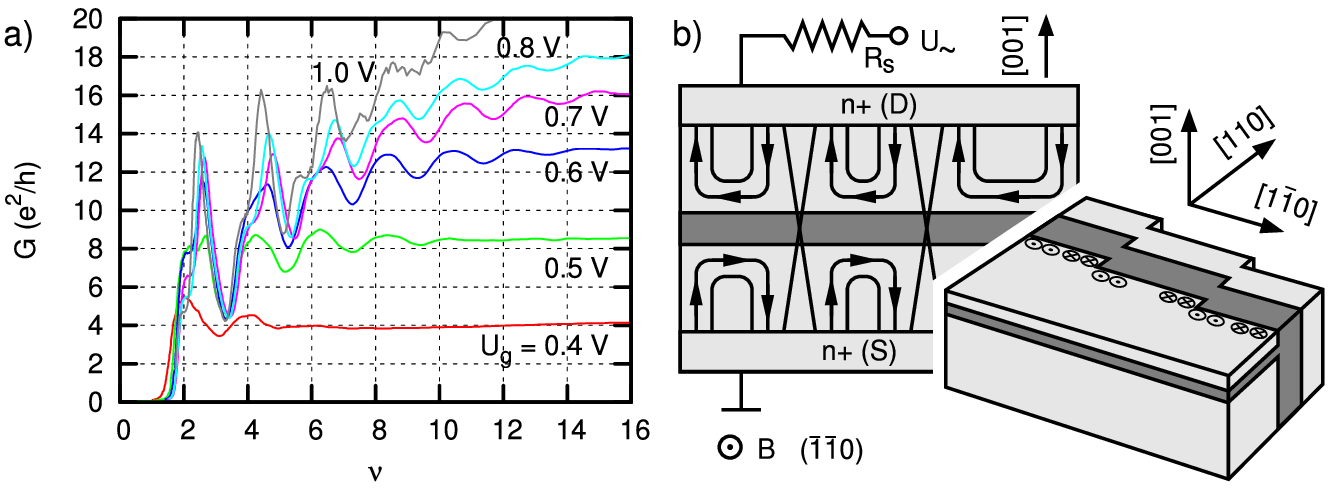}\vchcaption{\label{fig:enhanced}
  a) Landau oscillations from Fig.~\ref{fig:curves}b as a plot of the
  conductance $G=I/V$ vs the filling factor. The representation is based
  on the values of the electron density in Fig.~\ref{fig:density}b. b)~Sketches
  of a quantum region which is divided into independent sections due
  to macroscopic defects like steps on the cleavage plane. The
  entrance and exit points of the edge channels are denoted with
  $\otimes$ and $\odot$, respectively.}
\end{vchfigure}

The two-point conductance is plotted in Fig.~\ref{fig:enhanced}a as a
function of the filling factor. Since the electron transport takes
place in (coupled) one-dimensional edge channels, it is advantageous
to calibrate the ordinate in units of the conductance quantum
$G_\text{0}\equiv e^\text{2}/h$.  As stated before in Sec.~\ref{sec:tun},
one expects for the spin-degenerate system that the first conductance
peak is limited to~$\text2e^\text{2}/h$. However, the peaks at
$\nu\approx\text2$ in Fig.~\ref{fig:enhanced}a have an amplitude on
the order of $\text{10}e^\text{2}/h$.  In contrast, Kang\etal.\ report
on conductance oscillations which reach a height of only about
$\text{0.1}e^2/h$ \cite{Kan00a,Yan05,Yan04}. Although the origin of
the low current is not resolved, it is much more easier to comprehend
a peak amplitude which falls below the twofold conductance quantum
than if this threshold is exceeded. The tight-binding model of
Ref.~\cite{Tak00} shows that the conductance stays below $\text2e^2/h$
if the relative hopping amplitude is $\lesssim\text{0.2}$. Our result
$G>\text2G_\text{0}$ contradicts not only this conclusion, but also
the theory of Ref.~\cite{Kim03aKim03b} which predicts $G_t=Ke^2/h$
with $K<\text1$ for two spin-polarized quantum Hall systems. Thus, the
enhancement of the conductance in our CEO structure is probably based on
features which are not considered in both models.

Macroscopic defects of the cleavage plane are able to separate the
\val{500}{\micro m} long quantum region into several independent
sections. The extended \nplus-layers connect these multiple active
regions in parallel as illustrated in Fig.~\ref{fig:enhanced}b. The main reason
for the fragmentation of the quantum region is the local injury of the
upper $[\text1\bar{\text1}\text0]$-ridge during the sample fabrication
and preparation. The most likely origin are steps and corrugations
which emerge during the cleavage process. The thin electron system can
already be divided by steps which have a height on the order of the
2DES thickness, namely, about~\val{15}{nm}. Another source for a
partial damage of the electron systems consists in the manipulations during
the preparation process.  The disruption of the quantum region 
hardly appears if the contacts are fabricated by means of
photolithography on the cleavage plane as done by
Kang\etal.\ \cite{Yan04}. In contrast to extended contact layers as
used here, small contacts access just one or very few independent
sections of the active region.

\section{Landau oscillations}\label{sec:LL-osci}
The periodic coincidence of the Fermi level with the Landau band gaps
is responsible for the sequence of conductance peaks shown in
Fig.~\ref{fig:enhanced}a. By varying the electron density, it is
possible to measure the position of a certain anticrossing in
dependence of the magnetic field.  For this purpose, the conductance
traces~$G(B)$ have been recorded for different gate voltages. It is a
consequence of the unresolved spin-splitting that the distance of the
conductance peaks on the scale of the filling factor is about~2
instead of~1. From the Land\'e factor $g_0=-0.44$ of GaAs
\cite{Her77a} it follows for the ratio between the Zeeman and Landau
splitting 
\begin{equation}
  \frac{|g_0|\mu_BB}\hbar\cyclfreq=\frac12\frac\effm{m_e}|g_0|\approx\frac1{68}.
\end{equation}
The high electron density of the coupled 2DESs makes it necessary to
consider additionally the influence of electron-electron interactions
on the spin-splitting \cite{Dem93}.  The corresponding exchange energy
enhances the Land\'e factor which then has---depending on the actual
sample---a value between $g^*=3$ and 7 \cite{Ush90Dol97Hua02}.
Nevertheless, in similarity to Refs.\ \cite{Kan00a}
and~\cite{Gra04aHub05}, the conductance traces of our sample reveal no
clear spin-dependent features. According to the experimental results
presented by Kang\etal.\ in Fig.~1 of Ref.~\cite{Yan04}, it is by all
means possible that each electron system shows Shubnikov-de Haas
oscillations with a transition between spin-degenerate and -resolved
Landau levels while this feature is entirely absent in the
magnetotunneling data \cite{Hab06b}.

\begin{SCfigure}[4][t]
  \hskip.5in\includegraphics{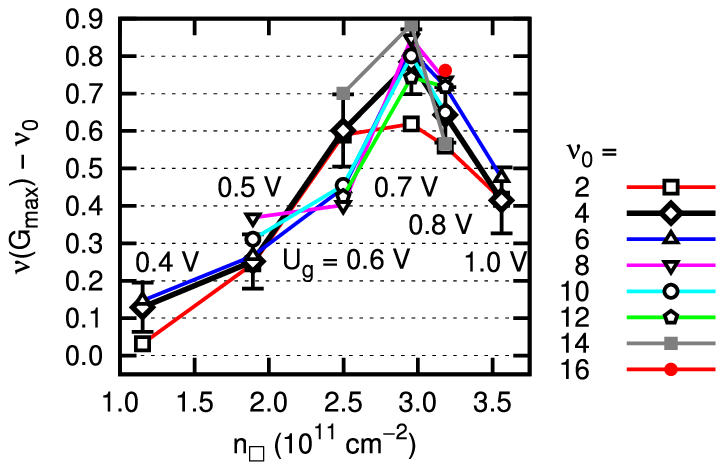}\caption{\label{fig:shift}Shift
  of the conductance peaks shown in Fig.~\ref{fig:enhanced}a in
  respect of a basis filling factor~$\nu_0$.  Error margins are shown
  for the $\nu_0=\text4$ peak.}
\end{SCfigure}

The expected magnetic shift of the anticrossings is small if one
considers the energy scale not absolutely, but in units of the
cyclotron energy. According to the band structure of
Fig.~\ref{fig:weak}b, the band gaps are shifted by
$\Delta\eps_B/\Delta B\approx\frac1{50}\frac1{\eval1T}$.  Likewise,
the conductance traces of Fig.~\ref{fig:enhanced}a have peaks with
positions which depend on the magnetic field. The peaks are shifted
for an increasing electron density towards higher filling factors. For
$\Ugate>\val{0.7}V$, they move to lower values again. In the
following, a basis filling factor $\nu_0=\text2$, 4, 6,~$\ldots$ is
assigned to the particular peaks of each conductance trace. The
difference between the actual peak position and $\nu_0$ is plotted in
Fig.~\ref{fig:shift}. For a constant basis filling factor~$\nu_0$, the
magnetic field increases in this graph from the left to the right.  In
order to keep the Fermi level within the same band gap, it is
necessary to increase the electron density for compensating an
enhanced degeneracy factor $s=eB/h$. The graphical comparison in
Fig.~\ref{fig:shift} shows that the conductance peaks evolve in a very
similar way for different basis filling factors.

The magnetic shift of the conductance peaks is compensated by another
effect which emerges for gate voltages greater than~\val{0.7}V.  The
corresponding decrease of the peak positions is not correlated with
the magnetic field, but is a consequence of the internal bias voltage
which builds up due to the gate leakage current.  The potential offset
between both electron films involves a lowering of the band gaps with
respect to the electron system next to the cathode \cite{Hab06b}.
This effect  shifts the conductance peaks towards lower filling
factors and has been revealed in Ref.~\cite{Kan00a} by applying an
external bias voltage. Our sample structure exhibits a very similar
behavior for the internal bias. The pronounced boundary
between rising and falling peak positions at $\Ugate=\val{0.7}V$ is
a consequence of the nearly exponential relation between the
gate and bias voltages (Fig.~\ref{fig:density}c).  Because the charge
carrier density of both electron systems changes simultaneously with
$\Vbias(\Ugate)$, the spectroscopy of a certain band gap for different
internal bias voltages requires an adjustment of the magnetic
field. Hence, the displacement of the conductance peaks cannot be
investigated independently from the magnetic shift of the Landau
gaps. Nevertheless, a quantitative comparison with the results of
Kang\etal., who applied an external bias, reveals similar peak
shifts. For a gate voltage which increases from 0.7 to \val{1.0}V, the
position of the  $\nu_0=\text4$ peak diminishes in Fig.~\ref{fig:shift}
by $\Delta\nu_B=-\text{0.36}$.  At the same time, the internal bias rises
according to Fig.~\ref{fig:density}b by
$\Delta\Vbias=\val{7.1}{mV}\approx \text{1.4}\hbar\cyclfreq/e$. For
this voltage offset, the first conductance peak in Fig.~2 of
Ref.~\cite{Kan00a} decreases from $\nu=\text{1.35}$ to~1.04. Thus, the
shift $\Delta\nu_B=-\text{0.31}$ obtained for an external bias is
comparable to the displacement encountered for an internal bias of
the same amplitude.

\begin{SCtable}[4][t]
  \hskip.5in\begin{tabular*}{.53\linewidth}{@{\extracolsep\fill}cccccc}
    \hline
    \rule[-1.2ex]{0pt}{3.8ex}\Ugate&$\nzwd/\text{10}^\text{11}$&
    $B$&$\nu(G_\text{max})$&\epsoben&\epsunten\\
    $(\text V)$&$(\text{cm}^{-\text2})$&$(\text T)$&&&\\[1pt]
    \hline
    0.4&1.16&1.15&4.13&2.062&2.072\\
    0.5&1.89&1.84&4.24&2.078&2.091\\
    0.6&2.49&2.25&4.58&2.086&2.100\\
    0.7&2.95&2.56&4.77&2.092&2.107\\
    0.8&3.18&2.84&4.63&2.097&2.113\\
  \hline
  \end{tabular*}
  \caption[]{\label{tab:cmp}Comparison between the position
    $\nu(G_\text{max})$ of the $\nu_0=\text4$ conductance peak and the
    corresponding band gap $[\epsoben;\epsunten]$ which separates
    the  Landau bands $\eps_{1k}$ and $\eps_{2k}$.}
\end{SCtable}

All conductance peaks in Fig.~\ref{fig:shift} are shifted towards
higher filling factors as the gate voltage increases up to~\val{0.7}V.
Due to the cropping of the conductance peak at $\nu_0=\text2$
(this feature will be discussed in Sec.~\ref{sec:AB}), the following
quantitative analysis is focused on the $\nu_0=\text4$ peak
which can be localized much more accurately. The band gap which is
responsible for this conductance peak is twofold, more precisely, it
consists of two anticrossings at opposed wave numbers
(Fig.~\ref{fig:weak}). Table~\ref{tab:cmp} compares the positions of
the $\nu_0=\text4$ conductance peak for different magnetic fields with
the result of corresponding band structure calculations. The relevant
band gap is characterized either by its extension
$[\epsoben;\epsunten]$ or its center $(\epsoben+\epsunten)/2$ and
width $(\epsunten-\epsoben)$.  For an increase of the gate voltage
from 0.4 to~\val{0.5}V, the magnetic field is raised from 1.15
to~\val{1.84}T in order to compensate for the enhanced electron
density and to keep the Fermi level between the Landau bands
$\eps_{1k}$ and $\eps_{2k}$. This shift corresponds to a displacement
of the conductance peak by $\Delta\nu_B^\text{exp}=\text{0.11}$ on the
scale of the filling factor.  According to the band structure
calculation, the higher magnetic field raises the energy gap by
$\Delta\eps_B^\text{theo}=\text{0.018}$.  The conversion to the
filling factor doubles this value: $\Delta\nu_B^\text{theo}=
\text{0.036}$.  For another data set, which is not shown here, we get
$\Delta\nu_B^\text{exp}=\text{0.06}$ and $2\Delta\eps_B^\text{theo}=
\text{0.026}$, respectively.

The ratio between the measured and predicted shift of the conductance
peak is $\Delta\nu_B^\text{exp}/ \Delta\nu_B^\text{theo}=
\text{2.7}\pm\text{0.4}$.  The discrepancy between both values is not
yet completely resolved, but it is probably a consequence of the
Coulomb interaction which is not considered in the calculated Landau
dispersion. The influence of this many body effect on the band gap
size is investigated in Refs.\ \cite{Mit01} and~\cite{Kol02}.  Both
author groups arrive at the result that the electron-electron
interaction enhances the energy gaps by a factor of about two. The
absolute position of the band gaps is, however, not explicitly
determined there.  Generally, the opposite edge channels are
converging with an increasing magnetic field according to
$\magl\propto B^{-\text{1/2}}$, with the consequence that the total
Coulomb energy rises. This effect is most likely responsible for the
encountered enhancement of the conductance peak displacement.

\section{Aharonov-Bohm effect}\label{sec:AB}
The Landau dispersion as discussed so far is valid for a perfect
tunneling barrier which is invariant in the $y$-direction. Within a
real barrier, however, there exist imperfections which lead to a local
enhancement of the tunneling probability.  These tunneling centers,
which are considered in the model of Kim and Fradkin
\cite{Kim03aKim03b}, give rise to Aharonov-Bohm (AB) interferences
between the opposite edge channels along the barrier
\cite{Yan05,Hab06a}.  In addition, the point contacts are responsible
for conductance fluctuations observed at low magnetic fields
\cite{Hab06b}. The latter effect which appears before the formation of
distinct quantum Hall edge channels will be discussed in
Sec.~\ref{sec:fluct_low}.

The rightmost resistance minimum of the curves in
Fig.~\ref{fig:curves}b shows some irregularities which are most
pronounced for $\Ugate=\val{0.5}V$ and disappear at higher gate
voltages. Figure~\ref{fig:max}a gives a magnification of this feature.
The left conductance peak is substantially cropped and the peak at
$\nu=\text{4.5}$ also exhibits slight irregularities. In general,
there exist two different types of oscillations which appear
simultaneously and exclusively at the first conductance peak on the
scale of the filling factor: short-period and quasiperiodic AB
oscillations as well as long-period conductance fluctuations. A visual
differentiation between both effects is possible by the black curves
in Figs. \ref{fig:max}a and~b. The two measurements differ only with
respect to the sweep velocity of the magnetic field. A high value for
$dB/dt$ in the right plot averages the AB oscillations out, whereas in
Fig.~\ref{fig:max}a both oscillation types are visible. The two kinds
of conductance fluctuations are not simply superimposed to the
long-range Landau oscillations, but they come along with a cropping
and flattening of the respective conductance peak.  While
Fig.~\ref{fig:max} reveals a clear distortion of the conductance peak at
\val{400}{mK}, the same maximum is in Fig.~1(a) of Ref.~\cite{Yan04}
for \val{300}{mK} entirely smooth. However, in Ref.~\cite{Yan05} also
a sample of Kang\etal.\ shows a strong cropping of the conductance peak
at a slightly lower temperature of \val{220}{mK}.

\begin{vchfigure}[t]
  \includegraphics{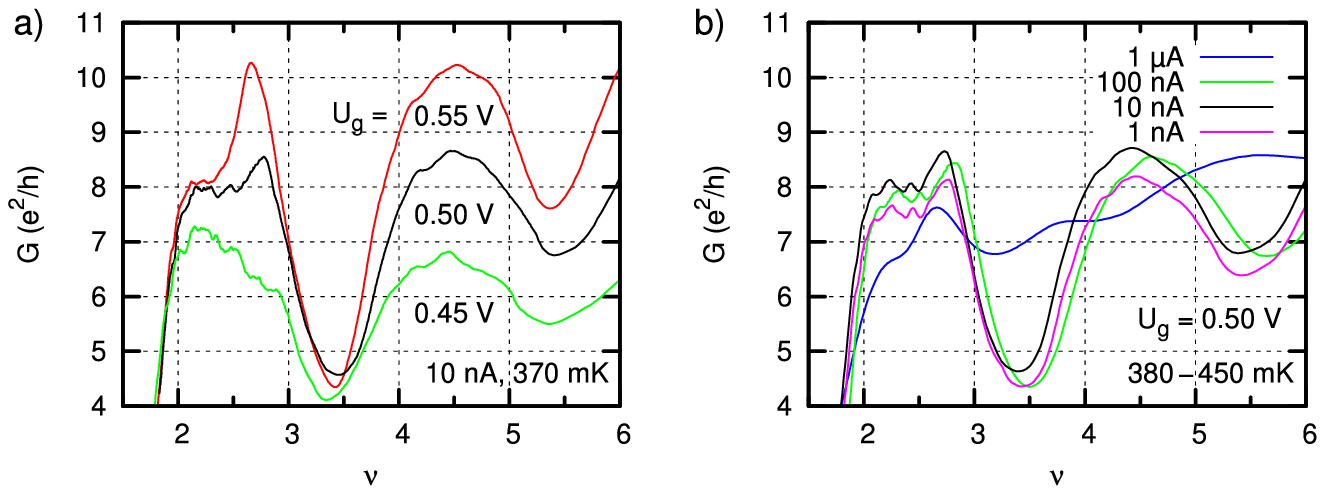}\vchcaption{\label{fig:max}a)
  Conductance traces for different gate voltages. b) Variation of the
  tunneling current at constant $\Ugate=\val{0.5}V$ and
  $U_\sim=\val1V$. The measurements have been carried out at an
  increased field sweep velocity.}
\end{vchfigure}

The phase coherence of the interfering electrons is harmed by both
thermal fluctuations and electrostatic potential gradients. The
critical temperature given by Eq.~\eqref{eqn:Tab} separates the
coherent and incoherent regime. The calculation of the band structure
for the mean magnetic field \val{3.5}T of the first conductance peak
yields $\langle x\rangle=\pm\text{1.2}\magl$ for  the
two involved opposite edge channels (inset of Fig.~\ref{fig:fft}b).
In anticipation of the mean distance of the tunneling centers,
$b=\val2{\micro m}$, the critical temperature is then
$T_\text{AB}=\val{130}{mK}$.  This result is insofar plausible as the
AB oscillations of Fig.~\ref{fig:max}a are almost suppressed at
\val{370}{mK} ($\Delta G/G\approx{0.2\percent}$) while in
Ref.~\cite{Yan05} they exhibit at \val{25}{mK} an amplitude which is
more easily recognizable ($\Delta G/G\approx\text{2\percent}$).

Each phase-destroying effect which appears in addition to thermal
fluctuations causes that some tunneling paths, first of all those with
point contacts at a great distance, lose their coherence.  Thereby,
the shape of the conductance fluctuations changes since they are
composed of all coherent AB oscillations.  The quantum interferences
are influenced by the internal voltage in a similar way as by thermal
fluctuations.  Due to the potential gradient across the barrier, a
tunneling electron possesses an excess energy with respect to the
Fermi level of the destination system. The bigger this difference, the
more free states are available around the final wave number, and the
more electron-electron scattering processes cause a loss of the phase
information \cite{Dat97}. Analogous to the thermal disorder, this
effect is negligible as long as
\begin{equation}
  e\Vbias\ll k_BT_\text{AB} \label{eqn:coh_Vbias}
\end{equation}
holds. The energy equivalent to the critical temperature is
$k_BT_\text{AB}=\val{11}{\micro eV}$. Consequently, the AB
oscillations are expected to be essentially undisturbed until the
internal bias voltage exceeds~\val{11}{\micro V}. This threshold is
reached in Fig.~\ref{fig:density}b for a gate voltage of
\val{0.498}V. Indeed, the conductance trace for $\Ugate=\val{0.50}V$
shows in Fig.~\ref{fig:max}a a significant indication of an initially
emerging incoherence: In comparison to the data for
$\Ugate=\val{0.45}V$, the right part of the conductance peak is
already partially smoothed for $\nu>2.6$. The effect is further
intensified at $\Ugate=\val{0.55}V$ in that the cropping of the peak
almost disappears while the AB signatures persist only at the curve
shoulder around $\nu=2.25$. Thus, the bias coherence condition is
essentially consistent with the experimental data as well as this is
the case for the temperature dependence.

\begin{vchfigure}
  \includegraphics{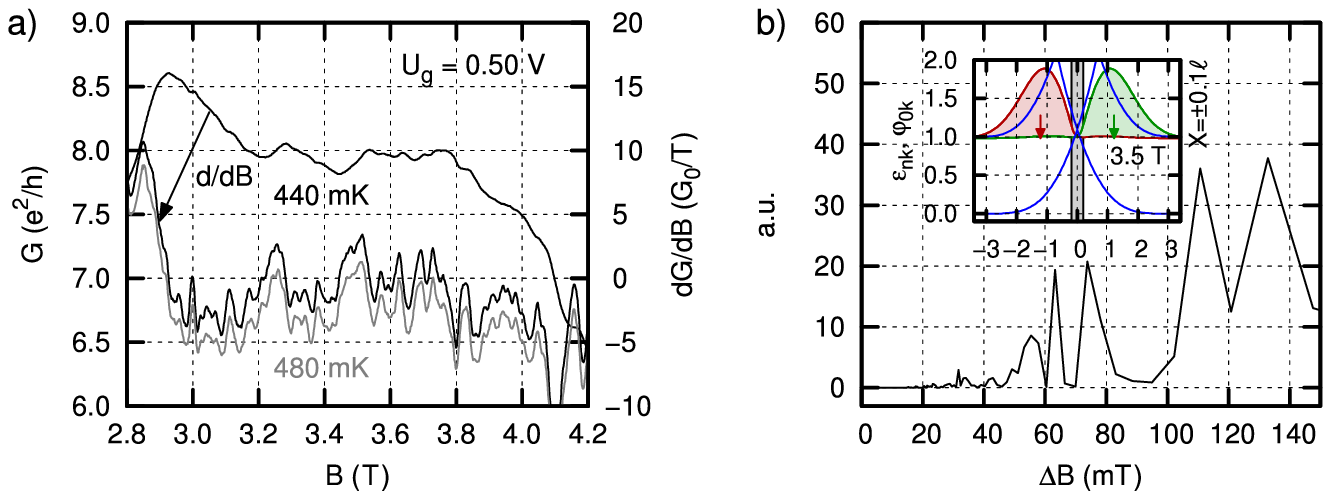}\vchcaption{\label{fig:fft}a)
  Magnification of the cropped conductance peak at
  $\nu=\text{2.5}$. The derivative is shown for two different field
  sweeps where the curve for~\val{480}{mK} is shifted downward for
  clarity. b)~Fast Fourier transform of $dG/dB$ for a field range of
  2.93--\val{4.26}T. Inset: the Landau dispersion (blue) and two wave
  functions are plotted vs $X$ and $x$, respectively. The shown energy
  eigenstates with $\eps_{0k}= \text{0.98}$ lie slightly below the
  first anticrossing at~1.09.  The arrows located at a distance of
  $a'=\text{2.4}\magl$ indicate the center of mass
  $\langle\varphi_{0k}|x|\varphi_{0k}\rangle$ of the left (red) and
  right (green) edge channel.}
\end{vchfigure}

Although the amplitude of the AB oscillations shown in
Fig.~\ref{fig:fft}a is rather small, they are clearly revealed in the
derivative of~$G(B)$. This is possible even for a high temperature of
about \val{500}{mK} which substantially exceeds the critical
temperature of $T_\text{AB}=\val{130}{mK}$.  The conductance traces,
which are composed of data points at a distance of \val{1.4}{mT}, are
smoothed before the differential quotient is calculated. For this
purpose, parabolas are fitted to the curve for each data point by
considering altogether 12 points in the proximity of the central
point. The slope of the parabola at the abscissa of the middle
sampling point is plotted in Fig.~\ref{fig:fft}a.  The comparison of
two successive field sweeps reveals that all fine structures of the
conductance traces are reproducible to a very high degree. It is
possible to resolve half cycles with a minimum width of~\val5{mT} and
an amplitude of at least~$\text{0.02}e^\text{2}/h$. The derivative
$dG/dB$ exhibits a strong quasiperiodic character which gives evidence
that there exist more than two tunneling centers within the
barrier. More precisely, this is an indication that several pairs of
point contacts simultaneously fulfill the coherence conditions of
Eqs.\ \eqref{eqn:coherence} and~\eqref{eqn:coh_Vbias}.  The Fourier
spectrum of Fig.~\ref{fig:fft}b is mainly concentrated within the two
period length intervals 52--\val{79}{mT}
and 106--\val{144}{mT}.  The comparison of the spectrum with the
conductance trace $G(B)$ in Fig.~\ref{fig:fft}a suggests that the
components of the second interval represent the second harmonic of the
long-period conductance fluctuations. The latter have a period of
$\Delta B\approx\val{0.23}T$ if the quantity is determined directly
from $G(B)$. The origin of these fluctuations will be discussed below.

According to Eq.~\eqref{eqn:AB_DeltaB}, the distance of two tunneling
centers which cause AB oscillations of a period~$\Delta B$ is
$b=h/\left(\Delta Ba'e\right)$. The distance~$a'$ of the involved
edge channels can be determined either by the approximation formula of
Eq.~\eqref{eqn:approx_distance} or from a dedicated band structure
calculation. The wave functions shown in the inset of
Fig.~\ref{fig:fft}b are located with their energy level next to the
first band gap and have a distance of~$\text{2.4}\magl$ if the
expectation value for the $x$-location is considered. For comparison,
Eq.~\eqref{eqn:approx_distance} yields for the two states with
$X=\pm\text{0.1}\magl$ an approximated distance
of~$\text{2.0}\magl$. Because the Landau dispersion varies only
slightly on the scales of the magnetic length and the cyclotron
energy, the exact result $a'=\text{2.4}\magl$ is also a good
approximation for system parameters which differ from those of
Fig.~\ref{fig:fft}.  However, in units of the barrier width, the
quantity~$a'$ may vary considerably: In our case ($a=\val{52}\AA$,
$B=\val{3.5}T$), the distance of the edge channels is $a'=\text{6.3}a$
while for the conditions of Ref.~\cite{Yan05} one gets
$a'=\text{3.0}a$. Generally, the distance of the opposite edge channels at
a thin barrier is less determined by the barrier width, but rather
by the extension $2R_c$ of the wave functions.  

The first interval of the Fourier spectrum of Fig.~\ref{fig:fft}b
contains contributions of prominent AB oscillations with periods of
$\Delta B=\text{56}$, 63, and \val{74}{mT}.  If the distance between
the interfering edge channels is $a'=\val{329}\AA$, these values
correspond to a distance of the tunneling centers of $b=\text{2.2}$,
2.0, and \val{1.7}{\micro m}, respectively.  However, as the
resolution of the Fourier spectrum is limited to about \val4{mT} due
to the finite width of the conductance peak, reliable conclusions
about the actual configuration of the tunneling centers are not
reasonable, so that we simply state the average distance of the point
contacts: $b=\eval{\left(2.0\pm0.2\right)}{\text{\micro}m}$.  If a
pair of tunneling centers with a distance of~$b$ belongs to the
coherent regime according to Eq.~\eqref{eqn:coherence}, the phase
coherence length is at least $L_\Phi^\text{AB}=a'+2b\simeq2b$ which
follows from Fig.~\ref{fig:tun}c.  Thus, our experimental data yields
$L_\Phi^\text{AB} \gtrsim\val{4.4}{\micro m}$.

The AB oscillations are superimposed by slowly varying conductance
fluctuations which have according to Fig.~\ref{fig:max}b a period
length of about \val{0.23}T. The authors of Ref.~\cite{Yan05} report a
similar period of about~\val{0.2}T. The long-period conductance
fluctuations occur in our samples for temperatures up
to~${\sim}\val{1.6}K$. Since this feature simultaneously emerges with
the AB oscillations and disappears together with them when the
internal bias increases (Fig.~\ref{fig:max}a), a connection to the
quantum interferences at tunneling centers is obvious.  However, the
theoretical work of Ref.~\cite{Kim03aKim03b} gives no direct evidence
for this effect.  Nevertheless, from microstructured AB rings it is
well-known that AB oscillations, which are periodic with respect to
the flux quantum~$\Phi_0$, simultaneously arise along with aperiodic
fluctuations which have a higher amplitude as well as a considerably
longer period \cite{Web85Web88,Umb84}.  The effect is attributed to
the finite width of an AB ring \cite{Sto85}. When the electrons
diffuse through the mesoscopic system, the corresponding trajectories
lie from time to time nearer to the inner or outer radius of the
ring-shaped electron path.  The corresponding fluctuations of the
enclosed magnetic flux exist in addition to the continuous variation
due to a rising or falling magnetic field. The random shape of the
electron trajectories within the conductor determines the phase
difference $2\pi\Phi/\Phi_0$ between both pathways and is reflected by
irregular fluctuations of the conductance. The effect can be reduced
or avoided if the area of the AB ring is very small compared to the
area  circumscribed by the whole ring path \cite{Sto85}.

In contrast to conventional AB rings, the interfering electrons in our
sample are not kept enclosed by a conductance path.  Their location is
rather determined by the magnetic confinement potential in combination
with the shape of the conductance band. For a varying magnetic field,
the enclosed flux $\Phi=Bba'$ changes by reason of two different
effects: Besides the proportionality with respect to the magnetic
field, the latter also controls the flux indirectly via the
distance~$a'$ of the interfering edge channels.  The field-dependence
of this length comes from the Landau dispersion and can be described
according to Eq.~\eqref{eqn:approx_distance} by the relation
$a'\approx\sqrt\nu\magl$.  The parameter~$a'$ is a smooth function of
the magnetic field only for an ideal system.  In a real sample, the
field-induced variation of the channel positions is additionally
influenced by the disorder potential. The spatial shift of the edge
channels is irregularly modified when the electrons at the Fermi edge
are forced by the disorder potential to propagate along a trajectory
which deviates from that expected for a perfect rectangular potential
wall. Because of $a'\ll b$ and $a'\sim\magl$, it is possible that
small displacements of the edge channels cause a variation of the
enclosed magnetic flux on the order of the flux quantum. However, due
to the local character of disorder, the effect emerges in comparison
to AB oscillations on considerably larger scales of the magnetic
field. The corresponding long-period and irregular conductance
oscillations represent an abstract image of the disorder at the
tunneling barrier.

\section{Conductance fluctuations at low magnetic fields}\label{sec:fluct_low}
The AB effect is restricted to the conductance peak which corresponds
to the lowest Landau band gap. Already at the subsequent peak, the
irregular oscillations are almost completely suppressed
(Fig.~\ref{fig:max}). Though the AB signatures rapidly disappear from
the conductance traces for a decreasing magnetic field, similar
fluctuations emerge at low field strengths again
(Fig.~\ref{fig:fluct}a). The quasiperiodic oscillations start from
zero field, get considerably stronger at about \val{0.27}T and vanish
at \val{0.8}T when the regular Landau oscillations arise. Just like
the AB oscillations, the conductance feature in the low field regime
is characterized by a quasiperiodic behavior and a high
reproducibility. Furthermore, the Fourier spectrum (inset of
Fig~\ref{fig:fluct}a) extends over a similar range of period, namely,
$\Delta B=\text{30}$--\val{65}{mT}.

\begin{vchfigure}[t]
  \includegraphics{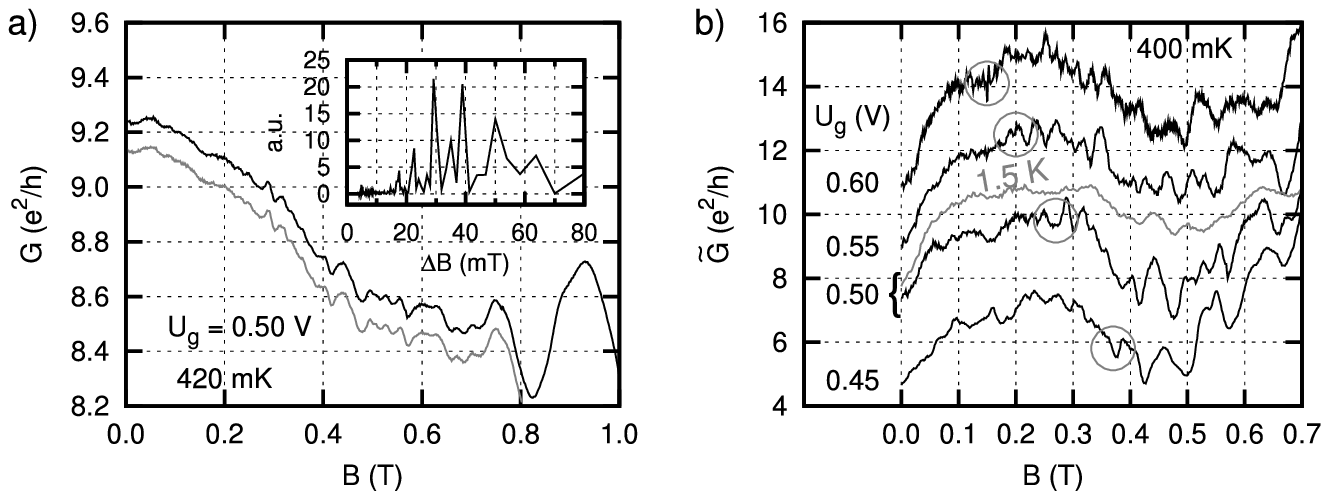}\vchcaption{\label{fig:fluct}a)
  Highly reproducible conductance fluctuations at small magnetic
  fields. The lower graph is shifted by $-\text{0.1}e^2/h$ for
  clarity. The inset contains the Fourier analysis of $G(B)$ for
  $B<\val{0.8}T$.  b)~Conductance traces for different gate
  voltages. For pointing out the details, each curve is plotted as the
  twentyfold difference between $G(B)$ and a straight line which is
  fitted to the experimental data. The circles mark the critical
  field~$B_c$ at which quasiperiodic oscillations with an amplitude of
  $\gtrsim\text{0.03}e^2/h$ become visible.}
\end{vchfigure}

In spite of the resemblance between the conductance fluctuations at
low magnetic fields and the quasiperiodic AB oscillations, there is
evidence that the low field feature is not based upon the interference
between counterpropagating edge channels.  The conductance traces of
Fig.~\ref{fig:fluct}b exhibit fluctuations which do not disappear for
an increasing gate voltage as it is the case for the AB oscillations
in Fig.~\ref{fig:max}a. This indicates that the quantum interferences
do not take place immediately within the internal bias gradient, but
they emerge apart from the tunneling barrier. Granted that the
conductance feature at low magnetic fields was caused by the quantum
interference of counterpropagating edge channels, one would expect
according to Eqs.~\eqref{eqn:AB_DeltaB}
and~\eqref{eqn:approx_distance} a period length of
\begin{equation}
  \Delta B\approx\frac1{b\sqrt\nu\,\magl}\,\frac he=
  \frac B{b\sqrt{\nzwd/2\pi}}. \label{eqn:deltaB_allg}
\end{equation}
In contrast to the AB oscillations at the $\nu_0=2$ peak, the
field-dependence of the period length cannot be neglected in the low
field regime. Though the conductance traces of Fig.~\ref{fig:fluct}b
fulfill $\Delta B\propto B$ at some degree, the absolute period of the
conductance fluctuations does not conform to the theoretical
prediction of Eq.~\eqref{eqn:deltaB_allg}: For the average distance
$b=\val{2.0}{\micro m}$ of the tunneling centers and the field range
0.27--\val{0.80}T this formula yields a period of 8--\val{23}{mT}
which substantially deviates from the Fourier spectrum of the
experimental data.

The conductance fluctuations at low magnetic fields originate most
likely from the bulk of the 2DESs. Recently, it has been shown with a
cooled scanning probe microscope that electrons which are injected
through a quantum point contact into a 2DES flow at zero magnetic
field along fan-shaped branches \cite{Top00,Aid07}. If just one
transversal mode passes the contact, the electrons are restricted to a
single branch. In our sample, the AlGaAs barrier contains tunneling
centers which are randomly distributed over the $W=\val{500}{\micro
m}$ long potential wall. From their average distance of~\val2{\micro
m}, one may estimate the number of tunneling centers to be on the
order of one hundred. In comparison, the 2DESs with the electron
density of \scival{1.89}{11}{cm$^{-\text2}$} at $\Ugate=\val{0.5}V$
possess $\lfloor k_FW/\pi\rfloor=\text{17244}$ transversal modes
\cite{Dat97}.  Even if the momentum relaxation length~$L_m$ fell below
the phase coherence length $L_\Phi^\text{AB}\gtrsim\val{4.4}{\micro
m}$ by two orders of magnitude, the localization length $L_c\approx
ML_m$ still clearly exceeds the extension $L_x=\val2{\micro m}$ of the
electron systems in the $x$-direction.  In addition, the short
electron systems represent a single phase coherence unit,
$L_x<L_\Phi^\text{AB}$, and, therefore, belong to the regime of weak
localization. However, this effect does generally not cause
conductance fluctuations in broad samples and would be anyway
destroyed by low magnetic fields \cite{Dat97}. As the imperfect
potential barrier reduces the mode number of the total system by at
least two orders of magnitude, strong localization ($L_x\approx L_c$)
becomes possible.  In this regime, the electron branches
which stem from the multiple slits in the tunneling barrier interfere
and cause conductance fluctuations while the Fermi level or the magnetic
field are varied. This effect vanishes at about~\val{0.8}T when all
electrons which are injected by the tunneling centers are forced by
the magnetic field to propagate within the same edge channels,
cf. Ref.~\cite{Aid07}, so that the fluctuating electron branches in
the bulk of the electron systems do not exist any longer.

The conductance fluctuations in the low field regime allow---similar
to the AB oscillations---an estimation of the phase coherence length. The
starting point is the semiclassical condition $B_c\mu_\Phi\gtrsim2\pi$
which requires electrons to fulfill at least one complete cyclotron
orbit.  The critical magnetic field~$B_c$ is determined by the onset
of significant oscillations which are defined here \emph{ad hoc} as a
`dense sequence' of fluctuations with an amplitude of minimally
$\text{0.03}e^\text{2}/h$.  For the conductance curve with
$\Ugate=\val{0.5}V$ in Fig.~\ref{fig:fluct}b, this definition yields a
phase coherence mobility of
$\mu_\Phi\approx\scival{2.3}5{cm$^\text{2}$/Vs}$. This quantity
corresponds to a phase coherence length of
$L_\Phi^{<\val1T}=\mu_\Phi\hbar k_F/e\approx \val{1.7}{\micro m}$
which is of the same order as the value
$L_\Phi^\text{AB}\gtrsim\val{4.4}{\micro m}$ obtained from the AB
oscillations.  For an increasing gate voltage, the momentum relaxation
time rises due to a more efficient screening of remote ionized
scatterers. The corresponding increase of the phase coherence length
$L_\Phi=v_F\sqrt{\tau_m\tau_\Phi/2}$ \cite{Dat97} is, indeed,
confirmed by a decrease of~$B_c$ in Fig.~\ref{fig:fluct}b.

\section{Summary}
We have presented a GaAs/AlGaAs heterostructure which contains two
laterally adjacent, field-induced quantum Hall systems which are
separated by a thin, epitaxially grown barrier. The structure includes
a gate electrode in order to allow a detailed analysis of the Landau
band structure at the line junction for different magnetic fields.
For the interpretation of the experimental data, the Landau dispersion
has been calculated exactly for non-interacting electrons residing in
a 2DES with a rectangular potential wall. The conductance traces in
dependence of the magnetic field are dominated by so-called Landau
oscillations which occur due to the periodic coincidence of the Fermi
level with the Landau band gaps. An observed enhancement of the
conductance which exceeds the predictions according to the
Landauer-B\"uttiker formalism is a consequence of macroscopic
defects. The latter divide the long and narrow quantum region into
multiple independent sections which are connected in parallel by the
extended \nplus-contact layers.  Our theory predicts for weakly
coupled electron systems a magnetic shift of the anticrossings on the
scale of the cyclotron energy of about
\val{1/50}{T$^{-\text{1}}$}. The experiment confirms this effect
semiquantitatively by a displacement of the conductance peaks on the
scale of the filling factor for an increasing Fermi level.

The conductance peak which corresponds to the gap between the first
two Landau bands is cropped and distorted by short- and long-period
conductance fluctuations.  The short-period feature is a result of the
AB interference between counterpropagating edge channels along the
barrier. The interference is made possible by tunneling centers which
exist accidentally within the AlGaAs barrier. For determining the
distance of the involved tunneling centers from the Fourier spectrum
of the conductance traces, it is necessary to know the distance of the
interfering edge channels.  By using the result of a band structure
calculation, we obtain an average distance of the point contacts of
about~\val2{\micro m}. The observed AB oscillations basically fulfill
two coherence conditions which describe the influence of the electron
temperature and the internal bias voltage, respectively.  The bias
results from gate leakage currents which lead to an offset between the
two coupled electron systems. Long-period conductance fluctuations at
the first conductance peak are a consequence of the disorder potential
which distorts the edge channel positions in an irregular way for a
varying magnetic field.

In addition to the interference effects at the cropped conductance
peak, we observe conductance fluctuations which emerge at small
magnetic fields and disappear at the onset of the regular Landau
oscillations. They occur in the regime of strong localization which is
a consequence of an imperfect barrier.  The latter represents a
multiple slit interferometer which essentially reduces the number of
transversal modes in the whole system. The phase coherence length
determined from the conductance fluctuations,
$L_\Phi^{<\val1T}\approx\val{1.7}{\micro m}$, is of the same order as
$L_\Phi^\text{AB}\gtrsim\val{4.4}{\micro m}$ which follows from the
distance of the tunneling centers involved for the AB oscillations.

\begin{acknowledgement}
This work was supported financially by the \emph{Deutsche
Forschungsgemeinschaft} (DFG) via the priority program
\emph{Quanten-Hall-Systeme}. We would like to thank M.~Bichler and
G.~Abstreiter as well as \hbox{M.~Reinwald} for sample growths. M.~H. is
grateful to M.~Grayson for useful and motivating discussions
throughout this project.
\end{acknowledgement}


\begin{thebibliography}{10}
\bibitem{Lan30} L. D. Landau, Z. Phys. \textbf{64}, 629 (1930).
\bibitem{Hal82} B. I. Halperin, Phys. Rev. B \textbf{25}, 2185 (1982).
\bibitem{Mac84} A. H. MacDonald and P. St\v{r}eda, Phys. Rev. B \textbf{29}, 
  1616 (1984). 
\bibitem{Kli80} K. v. Klitzing, G. Dorda, and M. Pepper, Phys. Rev. Lett.
  \textbf{45}, 494 (1980). 
\bibitem{Pfe90} L. Pfeiffer, K. W. West, H. L. Stormer, J. P. Eisenstein,
  K. W. Baldwin, D. Gershoni, and J. Spector, Appl. Phys. Lett. \textbf{56},
  1697 (1990). 
\bibitem{Gra04aHub05}M. Grayson, M. Huber, M. Rother, W. Biberacher,
  W. Wegscheider, M. Bichler, and G. Abstreiter, Physica E \textbf{25}, 212 (2004).\\
  M. Huber, M. Grayson, M. Rother, W. Biberacher, W. Wegscheider, and G. Abstreiter,
  Phys. Rev. Lett. \textbf{94}, 016805 (2005). 
\bibitem{Ho94} T.-L. Ho, Phys. Rev. B \textbf{50}, 4524 (1994).
\bibitem{Kan00a} W. Kang, H. L. Stormer, L. N. Pfeiffer, K. W. Baldwin, and K. W. West,
  Nature (Lond.) \textbf{403}, 59 (2000). 
\bibitem{Kim03aKim03b}E.-A. Kim and E. Fradkin, Phys. Rev. B \textbf{67},
  045317 (2003); 
  Phys. Rev. Lett. \textbf{91}, 156801 (2003). 
\bibitem{Yan05} I. Yang, W. Kang, L. N. Pfeiffer, K. W. Baldwin, K. W. West,
  E.-A. Kim, and E. Fradkin, Phys. Rev. B \textbf{71}, 113312 (2005).
\bibitem{Hab06a} M. Habl, M. Reinwald, W. Wegscheider, M. Bichler,
  and G. Abstreiter, Phys. Rev. B \textbf{73}, 205305 (2006).
\bibitem{Hab06b} M. Habl, Ph.D. thesis, Universit\"at Regensbug (2006).
\bibitem{Tak00} Y. Takagaki and K. H. Ploog, Phys. Rev. B \textbf{62},
  3766 (2000). 
\bibitem{HablCPC} The complete algorithm, including the consideration of
  the  effect of different effective masses in GaAs and AlGaAs, is to
  be published separately.
\bibitem{Sch04}R. Schuster, H. Hajak, M. Reinwald, W. Wegscheider, G. Schedelbeck, 
  S. Sedlmaier, M. Stopa, S. Birner, P.~Vogl, J. Bauer, D. Schuh, M. Bichler, and
  G. Abstreiter, phys. stat. sol. (c) \textbf{1}, 2028 (2004).
\bibitem{Fei06Her07}T. Feil, C. Gerl, and W. Wegscheider, Phys. Rev. B \textbf{73}, 
  125301 (2006).\\ 
  T. Herrle, Ph.D. thesis, Universit\"at Regensburg (2007).
\bibitem{Bar94}I. Barto\v s and B. Rosenstein, J.~Phys.~A: Math. Gen. \textbf{27}, 
  L53 (1994); 
  phys. stat. sol. (b) \textbf{193}, 411 (1996). 
\bibitem{Ada94}S. Adachi, GaAs and related materials (World Scientific, Singapore,
  1994).
\bibitem{Erd53Spa87}A. Erd\'elyi (ed.), Higher Transcendental Functions (McGraw-Hill,
  New York, 1953).\\
  J. Spanier and K. B. Oldham, An Atlas of Functions (Springer, Berlin, 1987).\\
  W. Magnus, F. Oberhettinger, and R. P. Soni, Formulas and Theorems for
  the Special Functions of Mathematical Physics (Springer, Berlin, 1966).
\bibitem{Lan65}L. D. Landau and E. M. Lifshitz, Quantum mechanics (Pergamon,
  Oxford, 1965).
\bibitem{Web85Web88}R. A. Webb, S. Washburn, C. P. Umbach, and R. B. Laibowitz,
  Phys. Rev. Lett. \textbf{54}, 2696 (1985).\\ 
  R. Webb and S. Washburn, Phys. Today~\textbf{41} (12), 46 (1988).
\bibitem{Yan04}I. Yang, W. Kang, K. W. Baldwin, L. N. Pfeiffer, and K. W. West,
  Phys. Rev. Lett. \textbf{92}, 056802 (2004). 
\bibitem{Her77a} C. Weisbuch and C. Hermann, Phys. Rev. B \textbf{15}, 816 (1977).
\bibitem{Dem93}J. Dempsey, B. Y. Gelfand, and B. I. Halperin, Phys. Rev. Lett. \textbf{70},
  3639 (1993). 
\bibitem{Ush90Dol97Hua02}A. Usher, R. J. Nicholas, J. J. Harris, and
  C. T. Foxon, Phys. Rev. B
  \textbf{41}, 1129 (1990).\\ 
  V. T. Dolgopolov, A. A. Shashkin, A. V. Aristov, D. Schmerek, W. Hansen, J. P. Kotthaus, and M. Holland,
  Phys. Rev. Lett. \textbf{79}, 729 (1997).\\
  T.-Y. Huang, Y.-M. Cheng, C.-T. Liang, G.-H. Kim, and J. Y. Leem,
  Physica~E \textbf{12}, 424 (2002). 
\bibitem{Mit01}A. Mitra and S. M. Girvin, Phys. Rev. B \textbf{64}, 041309 (2001).
\bibitem{Kol02}M. Kollar and S. Sachdev, Phys. Rev. B \textbf{65}, 121304 (2002).
\bibitem{Dat97}S. Datta, Electronic transport in mesoscopic systems 
  (Cambridge University Press, 1997). 
\bibitem{Umb84} C. P. Umbach, S. Washburn, R. B. Laibowitz, and R. A. Webb,
  Phys. Rev. B \textbf{30}, 4048 (1984).
\bibitem{Sto85} A. D. Stone, Phys. Rev. Lett. \textbf{54}, 2692 (1985).
\bibitem{Top00}M. A. Topinka, B. J. LeRoy, S. E. J. Shaw, E. J. Heller,
  R. M. Westervelt, K. D. Maranowski, and A. C. Gossard, Science \textbf{289},
  2323 (2000). 
\bibitem{Aid07}K. E. Aidala, R. E. Parrott, T. Kramer, E. J. Heller,
  R. M. Westervelt, M. P. Hanson, and A. C. Gossard, Nat. Phys. \textbf{3},
  464 (2007). 
\end{thebibliography}
\end{document}